\documentclass[journal]{IEEEtran}

\usepackage{cite}

\ifCLASSINFOpdf
\else
\fi

\usepackage[cmex10]{amsmath}

\usepackage{algorithmic}

\usepackage{amssymb}
\usepackage{bm}
\usepackage{amsfonts}

\usepackage{epsfig}

\usepackage{hyperref}

\usepackage{latexsym}
\usepackage{psfrag}
\usepackage{bm}
\usepackage{xspace}
\usepackage{graphicx}
\usepackage{subfigure}

\IEEEtriggeratref{59}

\renewcommand{\mathbf}[1]{{\bm{#1}}}     

\newcommand{\ignore}[1]{}

\newcommand{\supp}{\operatorname{supp}}

\newcommand{\dint}[1]{\operatorname{d}{#1}}

\newcommand{\matr}[1]{\mathbf{#1}}
\newcommand{\vect}[1]{\mathbf{#1}}
\newcommand{\code}[1]{\mathcal{#1}}

\newcommand{\set}[1]{\mathcal{#1}}
\newcommand{\graph}[1]{\mathsf{#1}}

\newcommand{\GF}[1]{\mathbb{F}_{#1}}
\newcommand{\R}{\mathbb{R}}

\newcommand{\Rp}{\mathbb{R}_{\geq 0}}
\newcommand{\Rpp}{\mathbb{R}_{> 0}}

\newcommand{\tr}{\mathsf{T}}

\newcommand{\codeCQC}[1]{\code{C}_{\mathrm{QC}}^{(r)}}

\newcommand{\defeq}{\triangleq}

\newcommand{\vnu}{\boldsymbol{\nu}}
\newcommand{\tvnu}{\boldsymbol{\tilde \nu}}
\newcommand{\vnuS}{\vnu_{\setS}}
\newcommand{\vnuSc}{\vnu_{\setSc}}
\newcommand{\vnnu}{\vnu}

\newcommand{\vnnuS}{\vnu_{\setS}}
\newcommand{\vnnuSc}{\vnu_{\setSc}}
\newcommand{\vlambda}{\boldsymbol{\lambda}}
\newcommand{\vomega}{\boldsymbol{\omega}}

\newcommand{\convhull}{\operatorname{conv}}
\newcommand{\conichull}{\operatorname{conic}}

\newcommand{\zeronorm}[1]{\lVert #1 \rVert_0}
\newcommand{\onenorm}[1]{\lVert #1 \rVert_1}
\newcommand{\twonorm}[1]{\lVert #1 \rVert_2}
\newcommand{\infnorm}[1]{\lVert #1 \rVert_{\infty}}
\newcommand{\zeroinfoperator}[1]{\lVert #1 \rVert_{0,\infty}}

\newcommand{\Absnorm}[1]{\left\lvert #1 \right\rvert}
\newcommand{\absnorm}[1]{\lvert #1 \rvert}
\newcommand{\Cnorm}[1]{\left\lvert #1 \right\rvert_{*}}

\newcommand{\Cvectnorm}[1]{\left\lVert #1 \right\rVert_{*}}

\newcommand{\norm}[2]{\lVert #1 \rVert_{#2}}

\newcommand{\vzero}{\vect{0}}
\newcommand{\vone}{\vect{1}}
\newcommand{\va}{\vect{a}}
\newcommand{\tva}{\tilde{\vect{a}}}
\newcommand{\vb}{\vect{b}}

\newcommand{\ve}{\vect{e}}
\newcommand{\tve}{\vect{\tilde e}}

\newcommand{\veS}{\vect{e}_{\setS}}
\newcommand{\veSc}{\vect{e}_{\setSc}}
\newcommand{\hve}{\vect{\hat e}}

\newcommand{\vs}{\vect{s}}
\newcommand{\tvs}{\vect{\tilde s}}

\newcommand{\vu}{\vect{u}}
\newcommand{\hvu}{\vect{\hat u}}

\newcommand{\vX}{\vect{X}}
\newcommand{\vY}{\vect{Y}}
\newcommand{\vx}{\vect{x}}
\newcommand{\tvx}{\vect{\tilde x}}

\newcommand{\vy}{\vect{y}}
\newcommand{\ovy}{\overline{\vect{y}}}
\newcommand{\oy}{\overline{y}}

\newcommand{\Z}{\mathbb{Z}}
\newcommand{\Zp}{\Z_{\geq 0}}
\newcommand{\Zpp}{\Z_{> 0}}

\newcommand{\hvx}{\vect{\hat x}}
\newcommand{\C}{{\mathbb C}}

\renewcommand{\leq}{\leqslant}
\renewcommand{\geq}{\geqslant}

\newcommand{\setA}{\set{A}}

\newcommand{\setI}{\set{I}}
\newcommand{\setJ}{\set{J}}

\newcommand{\setS}{\set{S}}
\newcommand{\setSc}{\overline{\setS}}

\newcommand{\nullspaceR}{\operatorname{Nullsp}_{\R}}
\newcommand{\nullspaceC}{\operatorname{Nullsp}_{\C}}
\newcommand{\nullspaceGFtwo}{\operatorname{Nullsp}_{\GF{2}}}
\newcommand{\NSPR}{\mathrm{NSP}^{\leq}_{\R}}
\newcommand{\SNSPR}{\mathrm{NSP}^{<}_{\R}}
\newcommand{\setweightR}[2]{\Sigma_{\R^{#1}}^{(#2)}} 
\newcommand{\setweightGFtwo}[2]{\Sigma_{\GF{2}^{#1}}^{(#2)}} 

\newcommand{\varphiM}{\boldsymbol{\varphi}_M}

\newcommand{\dc}{d_{\mathrm{c}}}
\newcommand{\dv}{d_{\mathrm{v}}}

\DeclareMathOperator*{\argmax}{arg\,max}

\newtheorem{lemma}{Lemma}
\newtheorem{theorem}[lemma]{Theorem}

\newtheorem{corollary}[lemma]{Corollary}

\newtheorem{PreDefinition}[lemma]{Definition}
  \newenvironment{definition}%
    {\begin{PreDefinition}}{\hfill$\square$\end{PreDefinition}}

\newtheorem{PreRemark}[lemma]{Remark}
  \newenvironment{remark}%
    {\begin{PreRemark}\upshape}{\hfill$\square$\end{PreRemark}}

\newtheorem{PreExample}[lemma]{Example}
  \newenvironment{example}%
    {\begin{PreExample}\upshape}{\hfill$\square$\end{PreExample}}

\newcommand{\rank}{\operatorname{rank}}
\newcommand{\fp}[1]{\set{#1}}

\newcommand{\fc}[1]{\set{#1}}

\newcommand{\wpsAWGNC}{w_{\mathrm{p}}^{\mathrm{AWGNC}}}
\newcommand{\wpsBSC}{w_{\mathrm{p}}^{\mathrm{BSC}}}
\newcommand{\wpsBSCmod}{w_{\mathrm{p}}^{\mathrm{BSC}'}}
\newcommand{\wpsBEC}{w_{\mathrm{p}}^{\mathrm{BEC}}}

\newcommand{\wpsAWGNCmin}{w_{\mathrm{p}}^{\mathrm{AWGNC,min}}}
\newcommand{\wpsBSCmin}{w_{\mathrm{p}}^{\mathrm{BSC,min}}}
\newcommand{\wpsBECmin}{w_{\mathrm{p}}^{\mathrm{BEC,min}}}

\newcommand{\wmaxfr}{w_{\mathrm{max-frac}}}
\newcommand{\wmaxfrmin}{w^{\mathrm{min}}_{\mathrm{max-frac}}}

\newcommand{\card}[1]{|#1|}

\newcommand{\wH}{w_{\mathrm{H}}}

\newcommand{\codeCCC}{\code{C}_{\mathrm{CC}}}
\newcommand{\matrGCC}{\matr{G}_{\mathrm{CC}}}
\newcommand{\matrHCC}{\matr{H}_{\mathrm{CC}}}
\newcommand{\tmatrHCC}{\matr{\tilde H}_{\mathrm{CC}}}
\newcommand{\setXCC}{\set{X}_{\mathrm{CC}}}
\newcommand{\setYCC}{\set{Y}_{\mathrm{CC}}}
\newcommand{\codeCCCdown}[1]{\code{C}_{\mathrm{CC},#1}}

\newcommand{\codeCCS}{\code{C}_{\mathrm{CS}}}
\newcommand{\matrGCS}{\matr{G}_{\mathrm{CS}}}
\newcommand{\matrHCS}{\matr{H}_{\mathrm{CS}}}
\newcommand{\tmatrHCS}{\matr{\tilde H}_{\mathrm{CS}}}
\newcommand{\setXCS}{\set{X}_{\mathrm{CS}}}
\newcommand{\setYCS}{\set{Y}_{\mathrm{CS}}}

\newcommand{\inGFtwo}{\text{(mod $2$)}}

\newcommand{\CSOPT}{\textbf{CS-OPT}}
\newcommand{\CSLPD}{\textbf{CS-LPD}}
\newcommand{\CSOPTver}[1]{\textbf{CS-OPT#1}}
\newcommand{\CSLPDver}[1]{\textbf{CS-LPD#1}}

\newcommand{\CSOPTzeroinf}{\textbf{CS-OPT}_{0,\infty}}

\newcommand{\CSRELzeroinf}{\textbf{CS-REL}_{0,\infty}}

\newcommand{\CCMLD}{\textbf{CC-MLD}}
\newcommand{\CCLPD}{\textbf{CC-LPD}}
\newcommand{\CCMLDver}[1]{\textbf{CC-MLD#1}}
\newcommand{\CCLPDver}[1]{\textbf{CC-LPD#1}}

\newcommand{\CCMLDone}{\textbf{CC-MLD1}}
\newcommand{\CCMLDtwo}{\textbf{CC-MLD2}}

\newcommand{\tgraph}[2]{\graph{#1}(\matr{#2})}

\newcommand{\eg}{\emph{e.g.}}
\newcommand{\ie}{\emph{i.e.}}
\newcommand{\etal}{\emph{et al.}}
\newcommand{\confer}{\emph{cf.}}

\newcommand{\optprog}[2]
{%
  \noindent\mbox{}\\[0cm]
  \noindent\fbox{%
  \begin{minipage}{0.955\linewidth}
    \mbox{}\\[-0.5cm]
    #1\\[#2]
  \end{minipage}
  }
  \noindent\mbox{}\\[-0.2cm]
}

\newcommand{\optprognoframe}[2]
{%
  \noindent\mbox{}\\[0cm]
  \noindent\mbox{%
  \begin{minipage}{0.955\linewidth}
    \mbox{}\\[-0.5cm]
    #1\\[#2]
  \end{minipage}
  }
  \noindent\mbox{}\\[-0.2cm]
}

\newcounter{mytempeqcounter}

\begin{document}

\title{LDPC Codes for Compressed Sensing}

\author{Alexandros G.~Dimakis,~\IEEEmembership{Member,~IEEE}, \ 
        Roxana Smarandache,~\IEEEmembership{Member,~IEEE}, and \\
        Pascal O.~Vontobel,~\IEEEmembership{Member,~IEEE}%
  \thanks{To appear in IEEE Transactions on Information Theory, 2012.
                  Submitted, 
                  December 1, 2010.
                  Revised,
                  November 23, 2011,
                  and
                  December 11, 2011.
          The second author
          was partially supported by NSF Grants DMS-0708033 and CCF-0830608. 
          Parts of this work were presented at the 
          47th Allerton Conference on Communications, Control, and Computing,
          Allerton House, Monticello, Illinois, USA, Sep.~30--Oct.~2, 
          2009~\cite{Dimakis:Vontobel:09:1},
          and at the 2010 International Zurich Seminar on Communications,
          Zurich, Switzerland, Mar.~3--5, 
          2010~\cite{Dimakis:Smarandache:Vontobel:10:1}.}%
  \thanks{A.~G.~Dimakis is with the Department of Electrical 
          Engineering-Systems, University of Southern California,
          Los Angeles, CA 90089, USA (e-mail: dimakis@usc.edu).}%
  \thanks{R.~Smarandache is with the Department of Mathematics and
          Statistics, San Diego State University, San Diego, CA 92182, USA
          (e-mail: rsmarand@sciences.sdsu.edu).}%
  \thanks{P.~O.~Vontobel is with Hewlett--Packard Laboratories, 1501
          Page Mill Road, Palo Alto, CA 94304, USA.  (e-mail:
          pascal.vontobel@ieee.org).}%
      }

\maketitle

\begin{abstract}
  We present a mathematical connection between channel coding and compressed
  sensing. In particular, we link, on the one hand, \emph{channel coding
    linear programming decoding (CC-LPD)}, which is a well-known relaxation of
  maximum-likelihood channel decoding for binary linear codes, and, on the
  other hand, \emph{compressed sensing linear programming decoding (CS-LPD)},
  also known as basis pursuit, which is a widely used linear programming
  relaxation for the problem of finding the sparsest solution of an
  under-determined system of linear equations. More specifically, we establish
  a tight connection between CS-LPD based on a zero-one measurement matrix
  over the reals and CC-LPD of the binary linear channel code that is obtained
  by viewing this measurement matrix as a binary parity-check matrix. This
  connection allows the translation of performance guarantees from one setup
  to the other. The main message of this paper is that parity-check matrices
  of ``good'' channel codes can be used as provably ``good'' measurement
  matrices under basis pursuit. In particular, we provide the first
  deterministic construction of compressed sensing measurement matrices with
  an order-optimal number of rows using high-girth low-density parity-check
  (LDPC) codes constructed by Gallager.
\end{abstract}

\begin{IEEEkeywords}
  Approximation guarantee,
  basis pursuit, 
  channel coding,
  compressed sensing,
  graph cover,
  linear programming decoding,
  pseudo-codeword,
  pseudo-weight,
  sparse approximation,
  zero-infinity operator.
\end{IEEEkeywords}

\IEEEpeerreviewmaketitle

\section{Introduction}
\label{sec:introduction}

\IEEEPARstart{R}{ecently}, there has been substantial interest in the theory
of recovering sparse approximations of signals that satisfy linear
measurements. Compressed sensing research (see, for example
\cite{Candes:Tao:05:1,DonohoCS}) has developed conditions for measurement
matrices under which (approximately) sparse signals can be recovered by
solving a linear programming relaxation of the original NP-hard combinatorial
problem. This linear programming relaxation is usually known as ``basis
pursuit.''

In particular, in one of the first papers in this area,
\confer~\cite{Candes:Tao:05:1}, Cand{\`e}s and Tao presented a setup they called
``decoding by linear programming,'' henceforth called compressed sensing
linear programming decoding (\CSLPD), where the sparse signal corresponds to
real-valued noise that is added to a real-valued signal that is to be
recovered in a hypothetical communication problem.

At about the same time, in an independent line of research, Feldman,
Wainwright, and Karger considered the problem of decoding a binary linear code
that is used for data communication over a binary-input memoryless channel, a
problem that is also NP-hard in general. In~\cite{Feldman:03:1,
  Feldman:Wainwright:Karger:05:1}, they formulated this channel coding problem
as an integer linear program, along with presenting a linear programming
relaxation for it, henceforth called channel coding linear programming
decoding (\CCLPD). Several theoretical results were subsequently proven about
the efficiency of \CCLPD, in particular for low-density parity-check (LDPC)
codes (see, \eg, \cite{Koetter:Vontobel:03:1, Vontobel:Koetter:05:1:subm,
  Feldman:Malkin:Servedio:Stein:Wainwright:04:1, DDKW07}).

As we will see in the subsequent sections, \CSLPD\ and \CCLPD\ (and the setups
they are derived from) look like similar linear programming relaxations,
however, a priori it is rather unclear if there is a connection beyond this
initial superficial similarity. The main technical difference is that \CSLPD\
is a relaxation of the objective function of a problem that is naturally over
the reals while \CCLPD\ involves a polytope relaxation of a problem defined
over a finite field. Indeed, Cand{\`e}s and Tao in their original paper asked
the question~\cite[Section VI.A]{Candes:Tao:05:1}: \emph{``\ldots In summary,
  there does not seem to be any explicit known connection with this line of
  work [\!\!\cite{Feldman:03:1, Feldman:Wainwright:Karger:05:1}] but it would
  perhaps be of future interest to explore if there is one.''}

In this paper we present such a connection between \CSLPD\ and \CCLPD. The
general form of our results is that if a given binary parity-check matrix is
``good'' for \CCLPD\ then the same matrix (considered over the reals) is a
``good'' measurement matrix for \CSLPD. The notion of a ``good'' parity-check
matrix depends on which channel we use (and a corresponding channel-dependent
quantity called pseudo-weight).

\begin{itemize}

\item Based on results for the binary symmetric channel (BSC), we show that if
  a parity-check matrix can correct any $k$ bit-flipping errors under \CCLPD,
  then the same matrix taken as a measurement matrix over the reals can be
  used to recover all $k$-sparse error signals under \CSLPD.

\item Based on results for binary-input output-symmetric channels with bounded
  log-likelihood ratios, we can extend the previous result to show that
  performance guarantees for \CCLPD\ for such channels can be translated into
  robust sparse-recovery guarantees in the $\ell_1 / \ell_1$ sense (see, \eg,
  \cite{GIKS_RIP}) for \CSLPD.

\item Performance guarantees for \CCLPD\ for the binary-input additive white
  Gaussian noise channel (AWGNC) can be translated into robust sparse-recovery
  guarantees in the $\ell_2 / \ell_1$ sense for \CSLPD.

\item Max-fractional weight performance guarantees for \CCLPD\ can be
  translated into robust sparse-recovery guarantees in the $\ell_{\infty} /
  \ell_1$ sense for \CSLPD.

\item Performance guarantees for \CCLPD\ for the binary erasure channel (BEC)
  can be translated into performance guarantees for the compressed sensing
  setup where the support of the error signal is known and the decoder tries
  to recover the sparse signal (\ie, tries to solve the linear equations) by
  back-substitution only.

\end{itemize}
All our results are also valid in a stronger, point-wise sense. For example,
for the BSC, if a parity-check matrix can recover a \emph{given set} of $k$
bit flips under \CCLPD, the same matrix will recover any sparse signal
supported on those $k$ coordinates under \CSLPD. In general, ``good''
performance of \CCLPD\ on a given error support set will yield ``good''
\CSLPD\ recovery for sparse signals supported by the same set.
 
It should be noted that all our results are only one-way: we do not prove that
a ``good'' zero-one measurement matrix will always be a ``good'' parity-check
matrix for a binary code. This remains an interesting open problem.

Besides these main results we also present reformulations of \CCLPD\ and
\CSLPD\ in terms of so-called graph covers: these reformulations will help in
seeing further similarities and differences between these two linear
programming relaxations. Moreover, based on an operator that we will call the
zero-infinity operator, we will define an optimization problem called
$\CSOPTzeroinf$, along with a relaxation of it called $\CSRELzeroinf$. Let
\CSOPT\ be the NP-hard combinatorial problem mentioned at the beginning of the
introduction whose relaxation is \CSLPD. First, we will show that
$\CSRELzeroinf$ is equivalent to $\CSLPD$. Secondly, we will argue that the
solution of \CSLPD\ is ``closer'' to the solution of $\CSOPTzeroinf$ than the
solution of \CSLPD\ is to the solution of \CSOPT. This is interesting because
$\CSOPTzeroinf$ is, like \CSOPT, in general an intractable optimization
problem, and so $\CSOPTzeroinf$ is at least as justifiably as \CSOPT\ a
difficult optimization problem whose solution is approximated by \CSLPD.

The organization of this paper is as follows. In Section~\ref{sec:notation:1}
we set up the notation that will be used. Then, in Sections~\ref{sec:cs:lpd:1}
and~\ref{sec:cc:lpd:1} we review the compressed sensing and channel coding
problems, along with their respective linear programming relaxations.

Section~\ref{sec:bridge:1} is the heart of this paper: it establishes the
lemma that will bridge \CSLPD\ and \CCLPD\ for zero-one matrices. Technically
speaking, this lemma shows that non-zero vectors in the real nullspace of a
measurement matrix (\ie, vectors that are problematic for \CSLPD) can be
mapped to non-zero vectors in the fundamental cone defined by that same matrix
(\ie, to vectors that are problematic for \CCLPD). 

Afterwards, in Section~\ref{sec:translation:1} we use the previously developed
machinery to establish the main results of this paper, namely the translation
of performance guarantees from channel coding to compressed sensing. By
relying on prior channel coding
results~\cite{Feldman:Malkin:Servedio:Stein:Wainwright:07:1, DDKW07,
  ADS:Improved_LP} and the above-mentioned lemma, we present novel results on
sparse compressed sensing matrices. Perhaps the most interesting corollary
involves the sparse deterministic matrices constructed in Gallager's
thesis~\cite[Appendix~C]{Gallager:63}. In particular, by combining our
translation results with a recent breakthrough by Arora
\etal~\cite{ADS:Improved_LP} we show that high-girth deterministic matrices
can be used for compressed sensing to recover sparse signals. To the best of
our knowledge, this is the first deterministic construction of measurement
matrices with an order-optimal number of rows.

Subsequently, Section~\ref{sec:reformulations:1} tightens the connection
between \CCLPD\ and \CSLPD\ with the help of graph covers, and
Section~\ref{sec:minimizing:zero:infty:norm:1} presents the above-mentioned
results involving the zero-infinity operator. Finally, some conclusions are
presented in Section~\ref{sec:conclusions:1}. 

The appendices contain the longer proofs. Moreover,
Appendix~\ref{sec:extensions:bridge:lemma:1} presents three generalizations of
the bridge lemma (\confer~Lemma~\ref{lemma:equation:nullspace:to:fc:1} in
Section~\ref{sec:bridge:1}) to certain types of integer and complex valued
matrices.

\section{Basic Notation}
\label{sec:notation:1}

Let $\Z$, $\Zp$, $\Zpp$, $\R$, $\Rp$, $\Rpp$, $\C$, and $\GF{2}$ be the ring
of integers, the set of non-negative integers, the set of positive integers,
the field of real numbers, the set of non-negative real numbers, the set of
positive real numbers, the field of complex numbers, and the finite field of
size $2$, respectively. Unless noted otherwise, expressions, equalities, and
inequalities will be over the field $\R$. The absolute value of a real number
$a$ will be denoted by $|a|$.

The size of a set $\setS$ will be denoted by $\card{\setS}$. For any $M \in
\Zpp$, we define the set $[M] \defeq \{ 1, \ldots, M \}$.

All vectors will be \emph{column} vectors. If $\vect{a}$ is some vector with
integer entries, then $\vect{a} \ (\mathrm{mod} \ 2)$ will denote an equally
long vector whose entries are reduced modulo $2$. If $\setS$ is a subset of
the set of coordinate indices of a vector $\va$ then $\va_{\setS}$ is the
vector with $\card{\setS}$ entries that contains only the coordinates of $\va$
whose coordinate index appears in $\setS$. Moreover, if $\va$ is a real vector
then we define $|\va|$ to be the real vector $\va'$ with the same number of
components as $\va$ and with entries $a'_i = |a_i|$ for all $i$. Finally, the
inner product $\langle \va, \vb \rangle$ of two equally long vectors $\va$ and
$\vb$ is written $\langle \va, \vb \rangle \defeq \sum_i a_i b_i$.

We define $\supp(\va) \defeq \{ i \ | \ a_i \neq 0 \}$ to be the support set
of some vector $\va$. Moreover, we let $\setweightR{n}{k} \defeq \bigl\{ \va
\in \R^n \bigm| \card{\supp(\va)} \leq k \bigr\}$ and $\setweightGFtwo{n}{k}
\defeq \bigl\{ \va \in \GF{2}^n \bigm| \card{\supp(\va)} \leq k \bigr\}$ be
the set of vectors in $\R^n$ and $\GF{2}^n$, respectively, which have at most
$k$ non-zero components. We refer to vectors in these sets as $k$-sparse
vectors.

For any real vector $\va$, we define $\zeronorm{\va}$ to be the $\ell_0$ norm
of $\va$, \ie, the number of non-zero components of $\va$. Note that
$\zeronorm{\va} = \wH(\va) = |\supp(\va)|$, where $\wH(\va)$ is the Hamming
weight of $\va$. Furthermore, $\onenorm{\va} \defeq \sum_i |a_i|$,
$\twonorm{\va} \defeq \sqrt{\sum_i |a_i|^2}$, and $\infnorm{\va} \defeq \max_i
|a_i|$ will denote, respectively, the $\ell_1$, $\ell_2$, and $\ell_{\infty}$
norms of $\va$.

For a matrix $\matr{M}$ over $\R$ with $n$ columns we denote its
$\R$-nullspace by $\nullspaceR(\matr{H}) \defeq \big\{ \va \in \R^n \bigm|
\matr{M} \cdot \va = \vzero \}$ and for a matrix $\matr{M}$ over $\GF{2}$ with
$n$ columns we denote its $\GF{2}$-nullspace by $\nullspaceGFtwo(\matr{H})
\defeq \big\{ \va \in \GF{2}^n \bigm| \matr{M} \cdot \va = \vzero \ \inGFtwo
\}$.

Let $\matr{H} = (h_{j,i})_{j,i}$ be some matrix. We denote the set of row and
column indices of $\matr{H}$ by $\setJ(\matr{H})$ and $\setI(\matr{H})$,
respectively. We will also use the sets $\setJ_i(\matr{H}) \defeq \{ j \in
\setJ \ | \ h_{j,i} \neq 0 \}$, $i \in \setI(\matr{H})$, and
$\setI_j(\matr{H}) \defeq \{ i \in \setI \ | \ h_{j,i} \neq 0 \}$, $j \in
\setJ(\matr{H})$. Moreover, for any set $\setS \subseteq \setI(\matr{H})$, we
will denote its complement with respect to $\setI(\matr{H})$ by $\setSc$, \ie,
$\setSc \defeq \setI(\matr{H}) \setminus \setS$. In the following, when no
confusion can arise, we will sometimes omit the argument $\matr{H}$ in the
preceding expressions.

Finally, for any $n, M \in \Zpp$ and any vector $\va \in \C^n$, we define the
$M$-fold lifting of $\va$ to be the vector $\va^{\uparrow M} = (a^{\uparrow
  M}_{(i,m)})_{(i,m)} \in \C^{Mn}$ with components given by
\begin{align*}
  a^{\uparrow M}_{(i,m)}
    &\defeq
       a_i,
       \quad (i,m) \in [n] \times [M].
\end{align*}
(One can think of $\va^{\uparrow M}$ as the Kronecker product of the vector
$\va$ with the all-one vector with $M$ components.) Moreover, for any vector
$\tva = (\tilde a_{(i,m)})_{(i,m)} \in \C^{Mn}$ or $\tva = (\tilde
a_{(i,m)})_{(i,m)} \in \GF{2}^{Mn}$ we define the projection of $\tva$ to the
space $\C^n$ to be the vector $\va \defeq \varphiM(\tva)$ with components
given by
\begin{align*}
  a_i
    &\defeq
       \frac{1}{M}
         \sum_{m \in [M]}
           \tilde a_{(i,m)},
        \quad i \in [n].
\end{align*}
(In the case where $\tva$ is over $\GF{2}$, the summation is over $\C$ and we
use the standard embedding of $\{ 0, 1 \}$ into $\C$.)

\section{Compressed Sensing \\ Linear Programming Decoding}
\label{sec:cs:lpd:1}

\subsection{The Setup}
\label{sec:cs:lpd:setup:1}

Let $\matrHCS$ be a real matrix of size $m \times n$, called the
\emph{measurement matrix}, and let $\vs$ be a real-valued vector containing
$m$ measurements. In its simplest form, the compressed sensing problem
consists of finding the sparsest real vector $\ve'$ with $n$ components that
satisfies $\matrHCS \cdot \ve' = \vs$, namely

\optprog
{
\begin{alignat*}{2}
  \CSOPT:
  \quad
  &
  \text{minimize} \quad
  &&
    \zeronorm{\ve'} \\
  &
  \text{subject to } \quad
  &&
   \matrHCS \cdot \ve' = \vs.
\end{alignat*}
}{-0.8cm}

\noindent
Assuming that there exists a sparse signal $\ve$ that satisfies the
measurement $\matrHCS \cdot \ve = \vs$, \CSOPT\ yields, for suitable matrices
$\matrHCS$, an estimate $\hve$ that equals $\ve$.

This problem can also be interpreted~\cite{Candes:Tao:05:1} as part of the
decoding problem that appears in a coded data communicating setup where the
channel input alphabet is $\setXCS \defeq \R$, the channel output alphabet is
$\setYCS \defeq \R$, and the information symbols are encoded with the help of
a real-valued code $\codeCCS$ of block length $n$ and dimension $\kappa \defeq
n - \rank_{\R}(\matrHCS)$ as follows.
\begin{itemize}

\item The code is $\codeCCS \defeq \bigl\{ \vx \in \R^n \bigm| \matrHCS \cdot
  \vx = \vzero \bigr\}$. Because of this, the measurement matrix $\matrHCS$ is
  sometimes also called an \emph{annihilator matrix}.

\item A matrix $\matrGCS \in \R^{n \times \kappa}$ for which $\codeCCS =
  \bigl\{ \matrGCS \cdot \vu \bigm| \vu \in \R^{\kappa} \bigr\}$ is
  called a \emph{generator matrix} for the code $\codeCCS$. With the help of
  such a matrix, information vectors $\vu \in \R^{\kappa}$ are encoded into
  codewords $\vx \in \R^n$ according to $\vx = \matrGCS \cdot \vu$.

\item Let $\vy \in \setYCS^n$ be the \emph{received vector}. We can write $\vy
  = \vx + \ve$ for a suitably defined vector $\ve \in \R^n$, which will be
  called the \emph{error vector}.  We initially assume that the channel is
  such that $\ve$ is \emph{sparse}, \ie, that the number of non-zero entries
  is bounded by some positive integer $k$. This will be generalized later to
  channels where the vector $\ve$ is \emph{approximately sparse}, \ie, where
  the number of large entries is bounded by some positive integer $k$.

\item The receiver first computes the syndrome vector $\vs$ according to $\vs
  \defeq \matrHCS \cdot \vy$. Note that
  \begin{align*}
    \vs
      &= \matrHCS \cdot (\vx + \ve)
       = \matrHCS \cdot \vx + \matrHCS \cdot \ve \\
      &= \matrHCS \cdot \ve.
  \end{align*}
  In a second step, the receiver solves \CSOPT\ to obtain an estimate $\hve$
  for $\ve$, which can be used to obtain the codeword estimate $\hvx = \vy -
  \hve$, which in turn can be used to obtain the information word estimate
  $\hvu$.
\end{itemize}

Because the complexity of solving \CSOPT\ is usually exponential in the
relevant parameters, one can try to formulate and solve a related optimization
problem with the aim that the related optimization problem yields very often
the same solution as \CSOPT, or at least very often a very good approximation
to the solution given by \CSOPT. In the context of \CSOPT, a popular approach
is to formulate and solve the following related optimization problem (which,
with the suitable introduction of auxiliary variables, can be turned into a
linear program):

\optprog
{
\begin{alignat*}{2}
  \CSLPD:
  \quad
  &
  \text{minimize} \quad
  &&
    \onenorm{\ve'} \\
  &
  \text{subject to } \quad
  &&
   \matrHCS \cdot \ve' = \vs.
\end{alignat*}
}{-0.8cm}

\noindent
This relaxation is also known as \emph{basis pursuit}.

\subsection{Conditions for the Equivalence of \CSLPD\  and \CSOPT}

A central question of compressed sensing theory is under what conditions the
solution given by \CSLPD\ equals (or is very close to) the solution given by
\CSOPT.\footnote{It is important to note that we worry only about the solution
  given by \CSLPD\ being equal (or very close) to the solution given by
  \CSOPT, because even \CSOPT\ might fail to correctly estimate the error
  vector in the above communication setup when the error vector has too many
  large components.}

Clearly, if $m \geq n$ and the matrix $\matrHCS$ has rank $n$, there is only
one feasible $\ve'$ and the two problems have the same solution.

In this paper we typically focus on the linear sparsity regime, \ie, $k =
\Theta(n)$ and $m = \Theta(n)$, but our techniques are more generally
applicable. The question is for which measurement matrices (hopefully with a
small number of measurements $m$) the LP relaxation is tight, \ie, the
estimate given by $\CSLPD$ equals the estimate given by $\CSOPT$. 

Celebrated compressed sensing results~(\eg
\cite{Candes-Romberg-Tao-06,DonohoCS}) established that ``good'' measurement
matrices exist. Here, by ``good'' measurement matrices we mean measurement
matrices that have only $m = \Theta\bigl( k \log (n/k) \bigr)$ rows and can
recover all (or almost all) $k$-sparse signals under \CSLPD. Note that for the
linear sparsity regime, $k = \Theta(n)$, the optimal scaling requires to
construct matrices with a number of measurements that scales linearly in the
signal dimension $n$.

One \emph{sufficient} way to certify that a given measurement matrix is
``good'' is the well-known restricted isometry property (RIP), indicating that
the matrix does not distort the $\ell_2$-norm of any $k$-sparse vector by too
much. If this is the case, the LP relaxation will be tight for all $k$-sparse
vectors $\ve$ and further the recovery will be robust to approximate
sparsity~\cite{Candes:Tao:05:1,Candes-Romberg-Tao-06,DonohoCS}. As is well
known, however, the RIP is not a complete characterization of the LP
relaxation of ``good'' measurement matrices (see, \eg,
\cite{Blanchard:Cartis:Tanner:11:1}). In this paper we use the nullspace
characterization instead (see, \eg, \cite{Xu:Hassibi:08:1,
  Stojnic:Xu:Hassibi:08:1}), that gives a necessary and sufficient condition
for a matrix to be ``good.''

\begin{definition}
  \label{def:nullspace:property:setS:1}

  Let $\setS \subseteq \setI(\matrHCS)$ and let $C \in \Rp$. We say that
  $\matrHCS$ has the nullspace property $\NSPR(\setS,C)$, and write $\matrHCS
  \in \NSPR(\setS,C)$, if
  \begin{align*}
    C \cdot \onenorm{\vnnuS}
      &\leq \onenorm{\vnnuSc},
       \quad
       \text{for all }
       \vnnu \in \nullspaceR(\matrHCS).
  \end{align*}
  We say that $\matrHCS$ has the strict nullspace property $\SNSPR(\setS,C)$,
  and write $\matrHCS \in \SNSPR(\setS,C)$, if
  \begin{align*}
    C \cdot \onenorm{\vnnuS}
      &< \onenorm{\vnnuSc},
       \quad
       \text{for all }
       \vnnu \in \nullspaceR(\matrHCS) \setminus \{ \vzero \}.\\[-0.75cm]
  \end{align*}
\end{definition}

\begin{definition}
  \label{def:nullspace:property:k:1}

  Let $k \in \Zp$ and let $C \in \Rp$. We say that $\matrHCS$ has the
  nullspace property $\NSPR(k,C)$, and write $\matrHCS \in \NSPR(k,C)$, if
  \begin{align*}
    \matrHCS
      &\in \NSPR(\setS,C),
       \ \ 
       \text{for all }
       \text{$\setS \subseteq \setI(\matrHCS)$
             with $\card{\setS} \leq k$}.
  \end{align*}
  We say that $\matrHCS$ has the strict nullspace property $\SNSPR(k,C)$, and
  write $\matrHCS \in \SNSPR(k,C)$, if
  \begin{align*}
    \matrHCS
      &\in \SNSPR(\setS,C),
       \ \ 
       \text{for all }
       \text{$\setS \subseteq \setI(\matrHCS)$
             with $\card{\setS} \leq k$}.\\[-0.75cm]
  \end{align*}
\end{definition}

Note that in the above two definitions, $C$ is usually chosen to be greater
than or equal to $1$.

As was shown independently by several authors (see \cite{Zhang_nullspace,
  Linial, FeuerCS, Stojnic:Xu:Hassibi:08:1} and references therein) the
nullspace condition in Definition~\ref{def:nullspace:property:k:1} is a
necessary and sufficient condition for a measurement matrix to be ``good'' for
$k$-sparse signals, \ie, that the estimate given by \CSLPD\ equals the
estimate given by \CSOPT\ for these matrices. In particular, the nullspace
characterization of ``good'' measurement matrices will be one of the keys to
linking \CSLPD\ with \CCLPD. Observe that the requirement is that vectors in
the nullspace of $\matrHCS$ have their $\ell_1$ mass spread in substantially
more than $k$ coordinates. (In fact, for $C \geq 1$, at least $2k$ coordinates
must be non-zero).

The following theorem is adapted from~\cite[Proposition~2]{FeuerCS}.

\begin{theorem}
  \label{theorem:sparse:error:1}

  Let $\matrHCS$ be a measurement matrix. Further, assume that $\vs = \matrHCS
  \cdot \ve$ and that $\ve$ has at most $k$ nonzero elements, \ie,
  $\zeronorm{\ve} \leq k$. Then the estimate $\hve$ produced by \CSLPD\ will
  equal the estimate $\hve$ produced by \CSOPT\ if $\matrHCS \in \SNSPR(k,
  C\!=\!1)$.
\end{theorem}

\noindent
\qquad \emph{Remark:} Actually, as discussed in~\cite{FeuerCS}, the condition
$\matrHCS \in \SNSPR(k, C\!=\!1)$ is also neces\-sary, but we will not use
this here.

The next performance metric (see, \eg, \cite{GIKS_RIP, CDD06}) for CS involves
recovering approximations to signals that are not exactly $k$-sparse.

\begin{definition}
  
  An $\ell_p/\ell_q$ approximation guarantee for \CSLPD\ means that
  \CSLPD\ outputs an estimate $\hve$ that is within a factor $C_{p,q}(k)$ from
  the best $k$-sparse approximation for $\ve$, \ie,
  \begin{align}
    \norm{\ve-\hve}{p}
      &\leq
        C_{p,q}(k)
        \cdot
        \min_{\ve' \in \setweightR{n}{k}}
          \norm{\ve-\ve'}{q},
            \label{eq:lp:lq;approximation:1}
  \end{align}
  where the left-hand side is measured in the $\ell_p$-norm and the right-hand
  side is measured in the $\ell_q$-norm.
\end{definition}

\noindent
Note that the minimizer of the right-hand side
of~\eqref{eq:lp:lq;approximation:1} (for any norm) is the vector $\ve' \in
\setweightR{n}{k}$ that has the $k$ largest (in magnitude) coordinates of
$\ve$, also called the best $k$-term approximation of
$\ve$~\cite{CDD06}. Therefore the right-hand side
of~\eqref{eq:lp:lq;approximation:1} equals $C_{p,q}(k) \cdot
\norm{\ve_{\overline{\setS^{*}}}}{q}$ where $\setS^{*}$ is the support set of
the $k$ largest (in magnitude) components of $\ve$. Also note that if $\ve$ is
$k$-sparse then the above condition suggests that $\hve = \ve$ since the right
hand-side of~\eqref{eq:lp:lq;approximation:1} vanishes, therefore it is a
strictly stronger statement than recovery of sparse signals. (Of course, such
a stronger approximation guarantee for $\hve$ is usually only obtained under
stronger assumptions on the measurement matrix.)

The nullspace condition is a necessary and sufficient condition on a
measurement matrix to obtain $\ell_1/\ell_1$ approximation guarantees. This is
stated and proven in the next theorem which is adapted
from~\cite[Theorem~1]{Xu:Hassibi:08:1}. (Actually, we omit the necessity part
in the next theorem since it will not be needed in this paper.)

\begin{theorem}
  \label{theorem:l1:l1:approximation:guarantee:1}

  Let $\matrHCS$ be a measurement matrix, and let $C > 1$ be a real
  constant. Further, assume that $\vs = \matrHCS \cdot \ve$. Then for any set
  $\setS \subseteq \setI(\matrHCS)$ with $\card{\setS} \leq k$ the solution
  $\hve$ produced by \CSLPD\ will satisfy
  \begin{align*}
    \onenorm{\ve - \hve}
      &\leq 2
            \cdot
            \frac{C+1}{C-1}
            \cdot
            \onenorm{\veSc}
  \end{align*}
  if $\matrHCS \in \NSPR(k,C)$.
\end{theorem}

\begin{IEEEproof}
  See Appendix~\ref{sec:proof:theorem:l1:l1:approximation:guarantee:1}.
\end{IEEEproof}

\section{Channel Coding \\ Linear Programming Decoding}
\label{sec:cc:lpd:1}

\subsection{The Setup}
\label{sec:cc:lpd:setup:1}

We consider coded data transmission over a memoryless channel with input
alphabet $\setXCC \defeq \{ 0, 1 \}$, output alphabet $\setYCC$, and channel
law $P_{Y | X}(y | x)$. The coding scheme will be based on a binary linear
code $\codeCCC$ of block length $n$ and dimension $\kappa$, $\kappa \leq
n$. In the following, we will identify $\setXCC$ with $\GF{2}$.
\begin{itemize}

\item Let $\matrGCC \in \GF{2}^{n \times \kappa}$ be a \emph{generator matrix}
  for $\codeCCC$. Consequently, $\matrGCC$ has rank $\kappa$ over $\GF{2}$,
  and information vectors $\vu \in \GF{2}^{\kappa}$ are encoded into codewords
  $\vx \in \GF{2}^n$ according to $\vx = \matrGCC \cdot \vu \ \inGFtwo$,
  \ie, $\codeCCC = \bigl\{ \matrGCC \cdot \vu \ \inGFtwo \bigm| \vu \in
  \GF{2}^{\kappa} \bigr\}$.\footnote{We remind the reader that throughout this
    paper we are using \emph{column} vectors, which is in contrast to the
    coding theory standard to use \emph{row} vectors.}

\item Let $\matrHCC \in \GF{2}^{m \times n}$ be a \emph{parity-check matrix}
  for $\codeCCC$. Consequently, $\matrHCC$ has rank $n-\kappa \leq m$ over
  $\GF{2}$, and any $\vx \in \GF{2}^n$ satisfies $\matrHCC \cdot \vx = \vzero
  \ \inGFtwo$ if and only if $\vx \in \codeCCC$, \ie, $\codeCCC = \bigl\{ \vx
  \in \GF{2}^n \bigm| \matrHCC \cdot \vx = \vzero \ \inGFtwo \bigr\}$.

\item In the following we will mainly consider the three following channels
  (see, for example, \cite{Richardson:Urbanke:08:1}): the binary-input
  additive white Gaussian noise channel (AWGNC, parameterized by its
  signal-to-noise ratio), the binary symmetric channel (BSC, parameterized by
  its cross-over probability), and the binary erasure channel (BEC,
  parameterized by its erasure probability).

\item Let $\vy \in \setYCC^n$ be the \emph{received vector} and define for
  each $i \in \setI(\matrHCC)$ the \emph{log-likelihood ratio} $\lambda_i
  \defeq \lambda_i(y_i) \defeq \log \bigl( \frac{P_{Y | X}(y_i|0)}{P_{Y |
      X}(y_i | 1)} \bigr)$.\footnote{On the side, let us remark that if
    $\setYCC$ is binary then $\setYCC$ can be identified with $\GF{2}$ and we
    can write $\vy = \vx + \ve \ \inGFtwo$ for a suitably defined vector $\ve
    \in \GF{2}^n$, which will be called the error vector. Moreover, we can
    define the \emph{syndrome vector} $\vs \defeq \matrHCC \cdot \vy \
    \inGFtwo$. Note that
    \begin{align*}
      \vs
        &= \matrHCC \cdot (\vx + \ve)
         = \matrHCC \cdot \vx + \matrHCC \cdot \ve \\
        &= \matrHCC \cdot \ve
           \quad \inGFtwo.
    \end{align*}
    However, in the following, with the exception of
    Section~\ref{sec:reformulations:1}, we will only use the log-likelihood
    ratio vector $\vlambda$, and not the binary syndrome vector $\vs$. (See
    Definition~\ref{def:syndrome:for:general:binary:input:channels:1} for a
    way to define a syndrome vector also for non-binary channel output
    alphabets $\setYCC$.)}

\end{itemize}
Upon observing $\vect{Y} = \vect{y}$, the \emph{(blockwise)
  maximum-likelihood decoding} (MLD) rule decides for
\begin{align*}
  \hat \vx(\vy)
    &= \argmax_{\vx' \in \codeCCC}
         P_{\vY | \vX}(\vy | \vx'),
\end{align*}
where $P_{\vY |\vX}(\vy | \vx') = \prod_{i \in \set{I}} P_{Y | X}(y_i |
x'_i)$. Formally:

\optprog
{
\begin{alignat*}{2}
  \CCMLD:
  \quad
  &
  \text{maximize} \quad
  &&
    P_{\vY |\vX}(\vy | \vx') \\
  &
  \text{subject to } \quad
  &&
   \vx' \in \codeCCC.
\end{alignat*}
}{-0.8cm}

\noindent It is clear that instead of $P_{\vY |\vX}(\vy | \vx')$ we can also
maximize $\log P_{\vY |\vX}(\vy | \vx') = \sum_{i \in \set{I}} \log P_{Y |
  X}(y_i | x'_i)$. Noting that $\log P_{Y | X}(y_i | x'_i) = - \lambda_i x'_i
+ \log P_{Y | X}(y_i | 0)$ for $x'_i \in \{ 0, 1 \}$, $\CCMLDone$ can then be
rewritten to read

\optprog
{
\begin{alignat*}{2}
  \CCMLDone:
  \quad
  &
  \text{minimize} \quad
  &&
    \langle \vlambda, \vx' \rangle \\
  &
  \text{subject to } \quad
  &&
   \vx' \in \codeCCC.
\end{alignat*}
}{-0.8cm}

\noindent Because the cost function is linear, and a linear function attains
its minimum at the extremal points of a convex set, this is
essentially equivalent to

\optprog
{
\begin{alignat*}{2}
  \CCMLDtwo:
  \quad
  &
  \text{minimize} \quad
  &&
    \langle \vlambda, \vx' \rangle \\
  &
  \text{subject to } \quad
  &&
   \vx' \in \convhull(\codeCCC).
\end{alignat*}
}{-0.8cm}

\noindent (Here, $\convhull(\codeCCC)$ denotes the convex hull of $\codeCCC$
after it has been embedded in $\R^n$. Note that we wrote ``essentially
equivalent'' because if more than one codeword in $\codeCCC$ is optimal for
$\CCMLDone$ then all points in the convex hull of these codewords are optimal
for $\CCMLDtwo$.) Although \CCMLDtwo\ is a linear program, it usually cannot
be solved efficiently because its description complexity is typically
exponential in the block length of the code.\footnote{Examples of code
  families that have sub-exponential description complexities in the block
  length are convolutional codes (with fixed state-space size), cycle codes
  (\ie, codes whose Tanner graph has only degree-$2$ vertices), and tree codes
  (\ie, codes whose Tanner graph is a tree). (For more on this topic, see for
  example~\cite{Kashyap:08:1}.) However, these classes of codes are not good
  enough for achieving performance close to channel capacity even under ML
  decoding (see, for example, \cite{Decreusefond:Zemor:94:1}.)}

However, one might try to solve a relaxation of \CCMLDtwo. Namely, as
proposed by Feldman, Wainwright, and Karger~\cite{Feldman:03:1,
  Feldman:Wainwright:Karger:05:1}, we can try to solve the optimization
problem

\optprog
{
\begin{alignat*}{2}
  \CCLPD:
  \quad
  &
  \text{minimize} \quad
  &&
    \langle \vlambda, \vx' \rangle \\
  &
  \text{subject to } \quad
  &&
   \vx' \in \fp{P}(\matrHCC),
\end{alignat*}
}{-0.8cm}

\noindent
where the relaxed set $\fp{P}(\matrHCC) \supseteq \convhull(\codeCCC)$ is
given in the next definition.

\begin{definition}
  For every $j \in \setJ(\matrHCC)$, let $\vect{h}^\tr_j$ be the $j$-th row of
  $\matrHCC$ and let
  \begin{align*}
    \codeCCCdown{j}
      &\defeq
         \big\{
           \vx \in \GF{2}^n
         \ \big| \ 
           \langle \vect{h}_j, \vx \rangle = 0
           \text{ (mod $2$)}
         \big\}.
  \end{align*}
  Then, the \emph{fundamental polytope} $\fp{P} \defeq \fp{P}(\matrHCC)$ of
  $\matrHCC$ is defined to be the set
  \begin{align*}
    \fp{P}
      &\defeq
    \fp{P}(\matrHCC)
       = \!\!\!\bigcap_{j \in \setJ(\matrHCC)}\!\!\!
           \convhull(\codeCCCdown{j}).
  \end{align*}
  Vectors in $\fp{P}(\matrHCC)$ will be called \emph{pseudo-codewords}.
\end{definition}

In order to motivate this choice of relaxation, note that the code $\codeCCC$
can be written as
\begin{align*}
  \codeCCC
    &= \codeCCCdown{1} \cap \cdots \cap \codeCCCdown{m},
\end{align*}
and so
\begin{align*}
  \convhull(\codeCCC)
    &= \convhull
         (
           \codeCCCdown{1} \cap \cdots \cap \codeCCCdown{m}
         ) \\
    &\subseteq
       \convhull
         (
           \codeCCCdown{1}) \cap \cdots \cap \convhull(\codeCCCdown{m}
         ) \\
    &= \fp{P}(\matrHCC).
\end{align*}
It can be verified~\cite{Feldman:03:1, Feldman:Wainwright:Karger:05:1} that
this relaxation possesses the important property that all the vertices of
$\convhull(\codeCCC)$ are also vertices of $\fp{P}(\matrHCC)$. Let us
emphasize that different parity-check matrices for the same code usually lead
to different fundamental polytopes and therefore to different \CCLPD{}s.

Similarly to the compressed sensing setup, we want to understand when we can
guarantee that the codeword estimate given by $\CCLPD$ equals the codeword
estimate given by $\CCMLD$.\footnote{It is important to note, as we did in the
  compressed sensing setup, that we worry mostly about the solution given by
  \CCLPD\ being equal to the solution given by \CCMLD, because even \CCMLD\
  might fail to correctly identify the codeword that was sent when the error
  vector is beyond the error correction capability of the code.} Clearly, the
performance of \CCMLD\ is a natural upper bound on the performance of \CCLPD,
and a way to assess \CCLPD\ is to study the gap to \CCMLD, \eg, by comparing
the here-discussed performance guarantees for \CCLPD\ with known performance
guarantees for \CCMLD.

When characterizing the \CCLPD\ performance of binary linear codes over
binary-input output-symmetric memoryless channels we can, without loss of
generality, assume that the all-zero codeword was
transmitted~\cite{Feldman:03:1, Feldman:Wainwright:Karger:05:1}. With this,
the success probability of \CCLPD\ is the probability that the all-zero
codeword yields the lowest cost function value when compared to all non-zero
vectors in the fundamental polytope. Because the cost function is linear, this
is equivalent to the statement that the success probability of \CCLPD\ equals
the probability that the all-zero codeword yields the lowest cost function
value compared to all non-zero vectors in the conic hull of the fundamental
polytope. This conic hull is called the \emph{fundamental cone} $\fc{K} \defeq
\fp{K}(\matrHCC)$ and it can be written as
\begin{align*}
  \fc{K}
    &\defeq
  \fc{K}(\matrHCC)
     = \conichull
         \big(
           \fp{P}(\matrHCC)
         \big)
     = \!\!\!\bigcap_{j \in \setJ(\matrHCC)}\!\!\!
         \conichull(\codeCCCdown{j}).
\end{align*}
The fundamental cone can be characterized by the inequalities listed in the
following lemma~\cite{Feldman:03:1, Feldman:Wainwright:Karger:05:1,
  Koetter:Vontobel:03:1, Vontobel:Koetter:05:1:subm,
  Koetter:Li:Vontobel:Walker:07:1}. (Similar inequalities can be given for the
fundamental polytope but we will not list them here since they are not needed
in this paper.)

\begin{lemma}
  \label{lemma:fundamental:cone:1}

  The fundamental cone $\fc{K} \defeq \fc{K}(\matrHCC)$ of $\matrHCC$ is the
  set of all vectors $\vomega \in \R^n$ that satisfy
  \begin{alignat}{2}
    \omega_i
      &\geq 0,
      \ \ \ 
      &&\text{for all $i \in \setI$,} 
          \label{eq:fund:cone:def:1} \\
    \omega_i
      &\leq
          \sum_{i' \in \setI_j \setminus i} \!\!
            \omega_{i'},
      \ \ \ 
      &&\text{for all $j \in \setJ$ 
              and all $i \in \setI_j$.}
          \label{eq:fund:cone:def:2}
  \end{alignat}
  \vskip-0.35cm
  \hfill$\square$
\end{lemma}

Note that in the following, not only vectors in the fundamental polytope, but
also vectors in the fundamental cone will be called
pseudo-codewords. Moreover, if $\matrHCS$ is a \emph{zero-one measurement
  matrix}, \ie, a measurement matrix where all entries are in $\{ 0, 1 \}$,
then we will consider $\matrHCS$ to represent also the parity-check matrix of
some linear code over $\GF{2}$. Consequently, its fundamental polytope will be
denoted by $\fp{P}(\matrHCS)$ and its fundamental cone by $\fc{K}(\matrHCS)$.

\subsection{Conditions for the Equivalence of \CCLPD\  and \CCMLD}

The following lemma gives a sufficient condition on $\matrHCC$ for \CCLPD\ to
succeed over a BSC.

\begin{lemma}
  \label{lemma:bsc:strict:balancedness:1}

  Let $\matrHCC$ be a parity-check matrix of a code $\codeCCC$ and let $\setS
  \subseteq \setI(\matrHCC)$ be the set of coordinate indices that are flipped
  by a BSC with non-zero cross-over probability. If $\matrHCC$ is such that
  \begin{align}
    \onenorm{\vomega_{\setS}}
        &< \onenorm{\vomega_{\setSc}}
             \label{eq:bsc:strict:balancedness:1}
  \end{align} 
  for all $\vomega \in \fc{K}(\matrHCC) \setminus \{ \vzero \}$, then the
  \CCLPD\ decision equals the codeword that was sent.
\end{lemma}

\noindent
\qquad \emph{Remark:} The above condition is also necessary; however, we will
not use this fact in the following.

\begin{IEEEproof}
  See Appendix~\ref{sec:proof:lemma:bsc:strict:balancedness:1}.
\end{IEEEproof}

\mbox{}

Note that the inequality in~\eqref{eq:bsc:strict:balancedness:1} is
\emph{identical} to the inequality that appears in the definition of the
strict nullspace property for $C = 1$\ (!). This observation makes one wonder
if there is a deeper connection between \CSLPD\ and \CCLPD\ beyond this
apparent one, in particular for measurement matrices that contain only zeros
and ones. Of course, in order to formalize a connection we first need to
understand how points in the nullspace of a zero-one measurement matrix
$\matrHCS$ can be associated with points in the fundamental polytope of the
parity-check matrix $\matrHCS$ (now seen as a parity-check matrix for a code
over $\GF{2}$). Such a mapping will be exhibited in the upcoming
Section~\ref{sec:bridge:1}. Before turning to that section, though, we need to
discuss pseudo-weights, which are a popular way of measuring the importance of
the different pseudo-codewords in the fundamental cone and which will be used
for establishing performance guarantees for \CCLPD.

\subsection{Definition of Pseudo-Weights}
\label{sec:pseudo:weight:definitions:1}

Note that the fundamental polytope and cone are functions only of the
parity-check matrix of the code and \emph{not} of the channel. The influence
of the channel is reflected in the pseudo-weight of the pseudo-codewords, so
it is only natural that every channel has its own pseudo-weight
definition. Therefore, every communication channel model comes with the right
measure of ``distance'' that determines how often a (fractional) vertex is
incorrectly chosen in \CCLPD.

\begin{definition}[\cite{Wiberg:96, Forney:Koetter:Kschischang:Reznik:01:1,
    Feldman:03:1, Feldman:Wainwright:Karger:05:1, Koetter:Vontobel:03:1,
    Vontobel:Koetter:05:1:subm}]
  \label{def:pseudo:weights:1}

  Let $\vomega$ be a non\-zero vector in $\Rp^n$ with $\vomega = (\omega_1,$
  $\ldots, \omega_n)$.
  \begin{itemize}
  
  \item The AWGNC pseudo-weight of $\vomega$ is defined to be
    \begin{align*}
      \wpsAWGNC(\vomega)
        &\defeq \frac{\onenorm{\vomega}^2}
                     {\twonorm{\vomega}^2}.
    \end{align*}

  \item In order to define the BSC pseudo-weight $\wpsBSC(\vomega)$, we let
    $\vomega'$ be the vector with the same components as $\vomega$ but in
    non-increasing order, \ie, $\vomega'$ is a ``sorted version'' of
    $\vomega$. Now let
    \begin{align*}
      f(\xi)
        &\defeq \omega'_i \quad (i-1 < \xi \leq i,\ 0 < \xi \leq n), \\
      F(\xi)
        &\defeq \int_{0}^{\xi} f(\xi') \dint{\xi'}, \\
      e
        &\defeq
           F^{-1} \left( \frac{F(n)}{2} \right)
         = F^{-1} \left( \frac{\onenorm{\vomega}}{2} \right).
    \end{align*}
    With this, the BSC pseudo-weight $\wpsBSC(\vomega)$ of $\vomega$ is
    defined to be $\wpsBSC(\vomega) \defeq 2e$.

  \item The BEC pseudo-weight of $\vomega$ is defined to be
    \begin{align*}
      \wpsBEC(\vomega)
        &= \big| \supp(\vomega) \big|.
    \end{align*}

  \item The max-fractional weight of $\vomega$ is defined to be
    \begin{align*}
      \wmaxfr(\vomega)
        &\defeq
           \frac{\onenorm{\vomega}}
                {\infnorm{\vomega}}.
    \end{align*}

  \end{itemize}
  For $\vomega = \vzero$ we define all of the above pseudo-weights and the
  max-fractional weight to be zero.\footnote{A detailed discussion of the
    motivation and significance of these definitions can be found
    in~\cite{Vontobel:Koetter:05:1:subm}.}
\end{definition}

For a parity-check matrix $\matrHCC$, the minimum AWGNC pseudo-weight is
defined to be
\begin{align*}
  \wpsAWGNCmin(\matrHCC)
    &\defeq
       \min_{\vomega \in \fp{P}(\matrHCC) \setminus \{ \vzero \}}
         \wpsAWGNC(\vomega) \\
    &= \min_{\vomega \in \fc{K}(\matrHCC) \setminus \{ \vzero \}}
         \wpsAWGNC(\vomega).
\end{align*}
The minimum BSC pseudo-weight $\wpsBSCmin(\matrHCC)$, the minimum BEC
pseudo-weight $\wpsBECmin(\matrHCC)$, and the minimum max-fractional weight
$\wmaxfrmin(\matrHCC)$ of $\matrHCC$ are defined analogously. Note that
although $\wmaxfrmin(\matrHCC)$ yields weaker performance guarantees than the
other quantities~\cite{Vontobel:Koetter:05:1:subm}, it has the advantage of
being efficiently computable~\cite{Feldman:03:1,
  Feldman:Wainwright:Karger:05:1}.

There are other possible definitions of a BSC pseudo-weight. For example, the
BSC pseudo-weight of $\vomega$ can also be taken to be
\begin{align*}
  \wpsBSCmod(\vomega)
    &\defeq
       \begin{cases}
         2e     & \text{if $\onenorm{\vomega'_{\{ 1, \ldots, e \}}}
                    = \onenorm{\vomega'_{\{ e+1, \ldots, n \}}}$} \\
         2e - 1 & \text{if $\onenorm{\vomega'_{\{ 1, \ldots, e \}}}
                    > \onenorm{\vomega'_{\{ e+1, \ldots, n \}}}$}
       \end{cases},
\end{align*}
where $\vomega'$ is defined as in Definition~\ref{def:pseudo:weights:1} and
where $e$ is the smallest integer such that $\onenorm{\vomega'_{\{ 1, \ldots,
    e \}}} \geq \onenorm{\vomega'_{\{ e+1, \ldots, n \}}}$. This definition of
the BSC pseudo-weight was for example used
in~\cite{Kelley:Sridhara:07:1}. (Note that
in~\cite{Forney:Koetter:Kschischang:Reznik:01:1} the quantity
$\wpsBSCmod(\vomega)$ was introduced as ``BSC effective weight.'')

Of course, the values $\wpsBSC(\vomega)$ and $\wpsBSCmod(\vomega)$ are tightly
connected. Namely, if $\wpsBSCmod(\vomega)$ is an even integer then
$\wpsBSCmod(\vomega) = \wpsBSC(\vomega)$, and if $\wpsBSCmod(\vomega)$ is an
odd integer then $\wpsBSCmod(\vomega) - 1 < \wpsBSC(\vomega) <
\wpsBSCmod(\vomega) + 1$.

The following lemma establishes a connection between BSC pseudo-weights and
the condition that appears in Lemma~\ref{lemma:bsc:strict:balancedness:1}.

\begin{lemma}
  \label{lemma:meaning:of:bsc:pseudo:weight:1}

  Let $\matrHCC$ be a parity-check matrix of a code $\codeCCC$ and let
  $\vomega$ be an arbitrary non-zero pseudo-codeword of $\matrHCC$, \ie,
  $\vomega \in \fc{K}(\matrHCC) \setminus \{ \vzero \}$. Then, for all sets
  $\setS \subseteq \setI(\matrHCC)$ with
  \begin{align*}
    \card{\setS}
      &< \frac{1}{2}
         \cdot
         \wpsBSC(\vomega) \ \text{ or with } \ 
    \card{\setS}
       < \frac{1}{2}
         \cdot
         \wpsBSCmod(\vomega),
  \end{align*}
  it holds that
  \begin{align*}
    \onenorm{\vomega_{\setS}}
      &< \onenorm{\vomega_{\setSc}}.
  \end{align*}
\end{lemma}

\begin{IEEEproof}
  See Appendix~\ref{sec:proof:lemma:meaning:of:bsc:pseudo:weight:1}.
\end{IEEEproof}

\section{Establishing a Bridge Between \\
               \CSLPD\  and \CCLPD}
\label{sec:bridge:1}

We are now ready to establish the promised bridge between \CSLPD\ and \CCLPD\
to be used in Section~\ref{sec:translation:1} to translate performance
guarantees from one setup to the other. Our main tool is a simple lemma that
was already established in~\cite{Smarandache:Vontobel:09:2}, but for a
different purpose. 

We remind the reader that we have extended the use of the absolute value
operator $\absnorm{\,\cdot\,}$ from scalars to vectors. So, if $\va = (a_i)_i$
is a real (complex) vector then we define $\absnorm{\va}$ to be the real
(complex) vector $\va' = (a'_i)_i$ with the same number of components as $\va$
and with entries $a'_i = \absnorm{a_i}$ for all $i$.

\begin{lemma}[Lemma~6 in \cite{Smarandache:Vontobel:09:2}]
  \label{lemma:equation:nullspace:to:fc:1}
 
  Let $\matrHCS$ be a zero-one measurement matrix. Then
  \begin{align*}
    \vnu \in \nullspaceR(\matrHCS)
    \ \ \ \Rightarrow \ \ \ 
    |\vnu| \in \fc{K}(\matrHCS).
  \end{align*}
\end{lemma}

\noindent
\qquad \emph{Remark:} Note that $\supp(\vnu) = \supp(|\vnu|)$.

\begin{IEEEproof}
  Let $\vomega \defeq |\vnu|$. In order to show that such a vector $\vomega$
  is indeed in the fundamental cone of $\matrHCS$, we need to
  verify~\eqref{eq:fund:cone:def:1} and~\eqref{eq:fund:cone:def:2}. The way
  $\vomega$ is defined, it is clear that it
  satisfies~\eqref{eq:fund:cone:def:1}. Therefore, let us focus on the proof
  that $\vomega$ satisfies~\eqref{eq:fund:cone:def:2}. Namely, from $\vnu \in
  \nullspaceR(\matrHCS)$ it follows that for all $j \in \setJ$, $\sum_{i \in
    \setI} h_{j,i} \nu_i = 0$, \ie, for all $j \in \setJ$, $\sum_{i \in
    \setI_j} \nu_i = 0$. This implies
  \begin{align*}
    \omega_i
      &= |\nu_i|
       = \left|
           \,\, - \!\!
           \sum_{i' \in \setI_j \setminus i}
             \nu_{i'}
         \right|
       \leq
         \sum_{i' \in \setI_j \setminus i}
            |\nu_{i'}|
       = \sum_{i' \in \setI_j \setminus i}
            \omega_{i'}
  \end{align*}
  for all $j \in \setJ$ and all $i \in \setI_j$, showing that $\vomega$
  indeed satisfies~\eqref{eq:fund:cone:def:2}.
\end{IEEEproof}

\mbox{}

This lemma gives a one-way result: with every point in the $\R$-nullspace of
the measurement matrix $\matrHCS$ we can associate a point in the fundamental
cone of $\matrHCS$, but not necessarily vice-versa. Therefore, a problematic
point for the $\R$-nullspace of $\matrHCS$ will translate to a problematic
point in the fundamental cone of $\matrHCS$ and hence to bad performance of
\CCLPD. Similarly, a ``good'' parity-check matrix $\matrHCS$ must have no low
pseudo-weight points in the fundamental cone, which means that there are no
problematic points in the $\R$-nullspace of $\matrHCS$. Therefore,
``positive'' results for channel coding will translate into ``positive''
results for compressed sensing, and ``negative'' results for compressed
sensing will translate into ``negative'' results for channel coding.

Further, Lemma~\ref{lemma:equation:nullspace:to:fc:1} preserves the support of
a given point $\vnu$. This means that if there are no low pseudo-weight points
in the fundamental cone of $\matrHCS$ with a given support, there are no
problematic points in the $\R$-nullspace of $\matrHCS$ with the same support,
which allows point-wise versions of all our results in
Section~\ref{sec:translation:1}.

Note that Lemma~\ref{lemma:equation:nullspace:to:fc:1} assumes that $\matrHCS$
is a zero-one measurement matrix, \ie, that it contains only zeros and
ones. As we show in Appendix~\ref{sec:extensions:bridge:lemma:1}, there are
suitable extensions of this lemma that put less restrictions on the
measurement matrix. However, apart from
Remark~\ref{remark:dense:measurement:matrix:1}, we will not use these
extensions in the following. (We leave it as an exercise to extend the results
in the upcoming sections to this more general class of measurement matrices.)

\section{Translation of Performance Guarantees}
\label{sec:translation:1}

In this section we use the above-established bridge between \CSLPD\ and
\CCLPD\ to translate ``positive'' results about \CCLPD\ to ``positive''
results about \CSLPD. Whereas Sections~\ref{sec:role:BSC:1}
to~\ref{sec:role:BEC:1} focus on the translation of abstract performance
bounds, Section~\ref{sec:translating:performance:guarantees:1} presents the
translation of numerical performance bounds. Finally, in
Section~\ref{sec:dense:measurement:matrices:1}, we briefly discuss some
limitations of our approach when dense measurement matrices are considered.

\subsection{The Role of the BSC Pseudo-Weight for \CSLPD}
\label{sec:role:BSC:1}

\begin{lemma}
  \label{lemma:from:bsc:to:cs:lpd:1}

  Let $\matrHCS \in \{ 0, 1 \}^{m \times n}$ be a CS measurement matrix and
  let $k$ be a non-negative integer. Then
  \begin{align*}
    \wpsBSCmin(\matrHCS) > 2k
    \ \ \ \Rightarrow \ \ \ 
    \matrHCS \in \SNSPR(k, C\!=\!1).
  \end{align*}
\end{lemma}

\begin{IEEEproof}
  Fix some $\vnu \in \nullspaceR(\matrHCS) \setminus \{ \vzero \}$. By
  Lemma~\ref{lemma:equation:nullspace:to:fc:1} we know that $|\vnu|$ is a
  pseudo-codeword of $\matrHCS$, and by the assumption $\wpsBSCmin(\matrHCS) >
  2k$ we know that $\wpsBSC(|\vnu|) > 2k$. Then, using
  Lemma~\ref{lemma:meaning:of:bsc:pseudo:weight:1}, we conclude that for all
  sets $\setS \subseteq \setI$ with $\card{\setS} \leq k$, we must have
  $\onenorm{\vnuS} = \onenorm{\,|\vnuS|\,} < \onenorm{\,|\vnuSc|\,} =
  \onenorm{\vnuSc}$. Because $\vnu$ was arbitrary, the claim $\matrHCS \in
  \SNSPR(k,C\!=\!1)$ clearly follows.
\end{IEEEproof}

\mbox{}

This result, along with Theorem~\ref{theorem:sparse:error:1} can be used to
establish sparse signal recovery guarantees for a compressed sensing matrix
$\matrHCS$.

Note that compressed sensing theory distinguishes between the so-called
\textbf{strong bounds} and the so-called \textbf{weak bounds}. The former bounds
correspond to a worst-case setup and guarantee the recovery of all $k$-sparse
signals, whereas the latter bounds correspond to an average-case setup and
guarantee the recovery of a signal on a randomly selected support with high
probability regardless of the values of the non-zero entries. Note that a
further notion of a weak bound can be defined if we randomize over the
non-zero entries also, but this is not considered in this
paper.

Similarly, for channel coding over the BSC, there is a distinction between
being able to recover from $k$ worst-case bit-flipping errors and being able
to recover from randomly positioned bit-flipping errors.

In particular, recent results on the performance analysis of \CCLPD\ have
shown that parity-check matrices constructed from expander graphs can correct
a constant fraction (of the block length $n$) of worst-case errors
(\confer~\cite{Feldman:Malkin:Servedio:Stein:Wainwright:07:1}) and random
errors (\confer~\cite{DDKW07, ADS:Improved_LP}). These worst-case error
performance guarantees implicitly show that the minimum BSC pseudo-weight of a
binary linear code defined by a Tanner graph with sufficient expansion
(expansion strictly larger than $3/4$) must grow linearly in $n$. (A
conclusion in a similar direction can be drawn for the random error setup.)
Now, with the help of Lemma~\ref{lemma:from:bsc:to:cs:lpd:1}, we can obtain
new performance guarantees for \CSLPD.

Let us mention that in~\cite{GIKS_RIP, XH_expander,
  Jafarpour:Xu:Hassibi:Calderbank:09:1}, expansion arguments were used to
directly obtain similar types of performance guarantees for compressed
sensing; in Section~\ref{sec:translating:performance:guarantees:1} we compare
these results to the guarantees we can obtain through our translation
techniques.

In contrast to the present subsection, which deals with the recovery of
(exactly) sparse signals, the next three subsections
(Sections~\ref{sec:beyond:BSC:1}, \ref{sec:role:AWGNC:1},
and~\ref{sec:role:maxfrac:weight:1}) deal with the recovery of approximately
sparse signals. Note that the type of guarantees presented in these
subsections are known as \textbf{instance optimality} guarantees~\cite{CDD06}.

\subsection{The Role of Binary-Input Channels Beyond the BSC
                     for \CSLPD}
\label{sec:beyond:BSC:1}

In Lemma~\ref{lemma:from:bsc:to:cs:lpd:1} we established a connection between,
on the one hand, performance guarantees for the BSC under \CCLPD, and, on the
other hand, the strict nullspace property $\SNSPR(k, C)$ for $C = 1$. It is
worthwhile to mention that one can also establish a connection between
performance guarantees for a certain class of binary-input channels under
\CSLPD\ and the strict nullspace property $\SNSPR(k, C)$ for $C > 1$. Without
going into details, this connection is established with the help of results
from~\cite{Feldman:Koetter:Vontobel:05:1}, that generalize results
from~\cite{Feldman:Malkin:Servedio:Stein:Wainwright:07:1}, and which deal with
a class of binary-input memoryless channels where all output symbols are such
that the magnitude of the corresponding log-likelihood ratio is bounded by
some constant $W \in \Rpp$.\footnote{Note that
  in~\cite{Feldman:Koetter:Vontobel:05:1}, ``This suggests that the asymptotic
  advantage over [\ldots] is gained not by quantization, but rather by
  restricting the LLRs to have finite support.''  should read ``This suggests
  that the asymptotic advantage over [\ldots] is gained not by quantization,
  but rather by restricting the LLRs to have bounded support.''} This
observation, along with Theorem~\ref{theorem:l1:l1:approximation:guarantee:1},
can be used to establish instance optimality $\ell_1 / \ell_1$ guarantees for
a compressed sensing matrix $\matrHCS$. Let us point out that in some recent
follow-up work~\cite{Khajehnejad:Tehrani:Dimakis:Hassibi:11:1} this has been
accomplished.

\subsection{Connection between AWGNC Pseudo-Weight and
                    $\ell_2 / \ell_1$ Guarantees}
\label{sec:role:AWGNC:1}

\begin{theorem}
  \label{theorem:weight:l2:l1:and:approximation:guarantees:1}

  Let $\matrHCS \in \{ 0, 1 \}^{m \times n}$ be a measurement matrix and let
  $\vs$ and $\ve$ be such that $\vs = \matrHCS \cdot \ve$. Let $\setS
  \subseteq \setI(\matrHCS)$ with $\card{\setS} = k$, and let $C'$ be an
  arbitrary positive real number with $C' > 4k$. Then the estimate $\hve$
  produced by \CSLPD\ will satisfy
  \begin{align*}
    \twonorm{\ve - \hve}
      &\leq \frac{C''}{\sqrt{k}}
            \cdot
            \onenorm{\veSc}
    \qquad
    \text{with}
    \qquad
    C''
      \defeq
        \frac{1}{\sqrt{\frac{C'}{4k}} - 1},
  \end{align*}
  if $\wpsAWGNC(|\vnnu|) \geq C'$ holds for all $\vnnu \in \nullspaceR(\matrHCS)
  \setminus \{ \vzero \}$. (In particular, this latter condition is satisfied
  for a measurement matrix $\matrHCS$ with $\wpsAWGNCmin(\matrHCS) \geq
  C'$.)
\end{theorem}

\begin{IEEEproof}
  See Appendix~\ref{sec:proof:theorem:weight:l2:l1:and:approximation:guarantees:1}.
\end{IEEEproof}

\subsection{Connection between Max-Fractional Weight and
                    $\ell_{\infty}/ \ell_1$ Guarantees}
\label{sec:role:maxfrac:weight:1}

\begin{theorem}
  \label{theorem:weight:linfty:l1:and:approximation:guarantees:1}

  Let $\matrHCS \in \{ 0, 1 \}^{m \times n}$ be a measurement matrix and let
  $\vs$ and $\ve$ be such that $\vs = \matrHCS \cdot \ve$. Let $\setS
  \subseteq \setI(\matrHCS)$ with $\card{\setS} = k$, and let $C'$ be an
  arbitrary positive real number with $C' > 2k$. Then the estimate $\hve$
  produced by \CSLPD\ will satisfy
  \begin{align*}
    \infnorm{\ve - \hve}
      &\leq \frac{C''}{k}
            \cdot
            \onenorm{\veSc}
    \qquad
    \text{with}
    \qquad
    C''
      \defeq
        \frac{1}{\frac{C'}{2k} - 1},
  \end{align*}
  if $\wmaxfr(|\vnnu|) {\geq} C'$ holds for all $\vnnu \in \nullspaceR(\matrHCS)
  \setminus \{ \vzero \}$. (In particular, this latter condition is satisfied
  for a measurement matrix $\matrHCS$ with $\wmaxfrmin(\matrHCS) \geq C'$.)
\end{theorem}

\begin{IEEEproof}
  See Appendix~\ref{sec:proof:theorem:weight:linfty:l1:and:approximation:guarantees:1}.
\end{IEEEproof}

\subsection{Connection between BEC Pseudo-Weight and \CSLPD}
\label{sec:role:BEC:1}

For the binary erasure channel, \CCLPD\ is identical to the peeling decoder
(see, \eg, \cite[Chapter~3.19]{Richardson:Urbanke:08:1}) that solves a system
of linear equations by only using back-substitution.

We can define an analogous compressed sensing problem by assuming that the
\emph{support} of the sparse signal $\ve$ is known to the decoder, and that
the recovering of the values is performed only by back-substitution. This
simple procedure is related to iterative algorithms that recover sparse
approximations more efficiently than by solving an optimization problem (see,
\eg, \cite{Tropp:Gilbert:07:1, Needell:Tropp:09:1, ZhangPfister_09, Montanari}
and references therein).

For this special case, it is clear that \CCLPD\ for the BEC and the described
compressed sensing decoder have identical performance since back-substitution
behaves exactly the same way over any field, be it the field of real numbers
or any finite field.  (Note that whereas the result of \CCLPD\ for the BEC
equals the result of the back-substitution-based decoder for the BEC, the same
is not true for compressed sensing, \ie, \CSLPD\ with given support of the
sparse signal can be strictly better than the back-substitution-based decoder
with given support of the sparse signal.)

\subsection{Explicit Performance Results}
\label{sec:translating:performance:guarantees:1}

In this section we use the bridge lemma,
Lemma~\ref{lemma:equation:nullspace:to:fc:1}, along with previous positive
performance results for \CCLPD, to establish performance results for the
\CSLPD~/ basis pursuit setup. In particular, three positive threshold results
for \CCLPD\ of low-density parity-check (LDPC) codes are used to obtain three
results that are, to the best of our knowledge, novel for compressed
sensing:
\begin{itemize}

\item \textbf{Corollary~\ref{cor:translation1}} (which relies on work by
  Feldman, Malkin, Servedio, Stein, and
  Wainwright~\cite{Feldman:Malkin:Servedio:Stein:Wainwright:07:1}) is very
  similar to~\cite{GIKS_RIP, XH_expander,
    Jafarpour:Xu:Hassibi:Calderbank:09:1}, although our proof is obtained
  through the connection to channel coding. We obtain a strong bound with
  similar expansion requirements.

\item \textbf{Corollary~\ref{cor:daskalakis:etal:1}} (which relies on work by
  Daskalakis, Dimakis, Karp, and Wainwright~\cite{DDKW07}) is a result that
  yields better constants (\ie, larger recoverable signals) but only with high
  probability over supports (\ie, it is a so-called weak bound).

\item \textbf{Corollary~\ref{cor:arora:etal:1}} (which relies on work by
  Arora, Daskala\-kis, and Steurer~\cite{ADS:Improved_LP}) is, in our opinion
  the most important contribution. We show the first deterministic
  construction of compressed sensing measurement matrices with an
  order-optimal number of measurements. Further we show that a property that
  is easy to check in polynomial time (\ie, girth), can be used to certify
  measurement matrices. Further, in the follow-up
  paper~\cite{Khajehnejad:Tehrani:Dimakis:Hassibi:11:1} it is shown that
  similar techniques can be used to construct the first optimal measurement
  matrices with $\ell_1 / \ell_1$ sparse approximation properties.

\end{itemize}
At the end of the section we also use
Lemma~\ref{lemma:equation:nullspace:to:fc:1:complex:case}
(\confer~Appendix~\ref{sec:extensions:bridge:lemma:1}) with $\Cnorm{\,\cdot\,}
= \absnorm{\,\cdot\,}$ to study dense measurement matrices with entries in $\{
-1, 0, +1 \}$.

Before we can state our first translation result, we need to introduce some
notation.

\begin{definition}
  Let $\graph{G}$ be a bipartite graph where the nodes in the two node classes
  are called left-nodes and right-nodes, respectively. If $\setS$ is some
  subset of left-nodes, we let $\set{N}(\setS)$ be the subset of the
  right-nodes that are adjacent to $\setS$. Then, given parameters $\dv \in
  \Zpp$, $\gamma \in (0,1)$, $\delta \in (0,1)$, we say that $\graph{G}$ is a
  $(\dv, \gamma, \delta)$-expander if all left-nodes of $\graph{G}$ have
  degree $\dv$ and if for all left-node subsets $\setS$ with $\card{\setS}
  \leq \gamma \cdot \card{\{\mathrm{left{-}nodes}\}}$ it holds that
  $\card{\set{N}(\setS)} \geq \delta \dv \cdot \card{\setS}$.
\end{definition}

\noindent Expander graphs have been studied extensively in past work on
channel coding (see, \eg, \cite{Sipser:Spielman:96}) and compressed sensing
(see, \eg, \cite{XH_expander, Jafarpour:Xu:Hassibi:Calderbank:09:1}). It is
well known that randomly constructed left-regular bipartite graphs are
expanders with high probability (see, \eg,
\cite{Feldman:Malkin:Servedio:Stein:Wainwright:07:1}).

In the following, similar to the way a Tanner graph is associated with a
parity-check matrix~\cite{Tanner:81}, we will associate a Tanner graph with a
measurement matrix. Note that the variable and constraint nodes of a Tanner
graph will be called left-nodes and right-nodes, respectively.

With this, we are ready to present the first translation result, which is a
so-called strong bound (\confer~the discussion in
Section~\ref{sec:role:BSC:1}). It is based on a theorem
from~\cite{Feldman:Malkin:Servedio:Stein:Wainwright:07:1}.

\begin{corollary}
  \label{cor:translation1}

  Let $\dv \in \Zpp$ and $\gamma \in (0,1)$. Let $\matrHCS \in \{ 0, 1 \}^{m
    \times n}$ be a measurement matrix such that the Tanner graph of
  $\matrHCS$ is a $(\dv, \gamma, \delta)$-expander with sufficient expansion,
  more precisely, with
  \begin{align*}
    \delta
      &> \frac{2}{3} +\frac{1}{3 \dv}
  \end{align*}
  (along with the technical condition $\delta \dv \in \Zpp$). Then \CSLPD\
  based on the measurement matrix $\matrHCS$ can recover all $k$-sparse
  vectors, \ie, all vectors whose support size is at most $k$, for
  \begin{align*}
    k
      &< \frac{3\delta - 2}
              {2\delta - 1}
         \cdot
         (\gamma n - 1). 
  \end{align*}
\end{corollary}

\begin{IEEEproof}
  This result is easily obtained by combining
  Lemma~\ref{lemma:equation:nullspace:to:fc:1}
  with~\cite[Theorem~1]{Feldman:Malkin:Servedio:Stein:Wainwright:07:1}.
\end{IEEEproof}

\mbox{}

Interestingly, for $\delta = 3/4$ the recoverable sparsity $k$ matches exactly
the performance of the fast compressed sensing algorithm in~\cite{XH_expander,
  Jafarpour:Xu:Hassibi:Calderbank:09:1} and the performance of the simple
bit-flipping channel decoder of Sipser an Spielman~\cite{Sipser:Spielman:96},
however, our result holds for the \CSLPD~/ basis pursuit setup. Moreover,
using results about expander graphs
from~\cite{Feldman:Malkin:Servedio:Stein:Wainwright:07:1}, the above corollary
implies, for example, that, for $m / n = 1/2$ and $\dv = 32$, sparse
expander-based zero-one measurement matrices will recover all $k = \alpha n$
sparse vectors for $\alpha \leq 0.000175$. To the best of our knowledge, the
only previously known result for sparse measurement matrices under basis
pursuit is the work of Berinde \etal~\cite{GIKS_RIP}. As shown by the authors
of that paper, the adjacency matrices of expander graphs (for expansion
$\delta > 5/6$) will recover all $k$-sparse signals. Further, these authors
also state results giving $\ell_1/\ell_1$ instance optimality sparse
approximation guarantees.  Their proof is directly done for the compressed
sensing problem and is therefore fundamentally different from our approach
which uses the connection to channel coding. The result of
Corollary~\ref{cor:translation1} implies a strong bound for all $k$-sparse
signals under basis pursuit and zero-one measurement matrices based on
expander graphs.  Since we only require expansion $\delta>3/4$, however, we
can obtain slightly better constants than~\cite{GIKS_RIP}.  Even though we
present the result of recovering exactly $k$-sparse signals, the results
of~\cite{Feldman:Koetter:Vontobel:05:1} can be used to establish
$\ell_1/\ell_1$ sparse recovery for the same constants.  We note that in the
linear sparsity regime $k=\alpha n$, the scaling of $m = c n$ is order optimal
and also the obtained constants are the best known for strong bounds of basis
pursuit. Still, these theoretical bounds are quite far from the observed
experimental performance. Also note that the work by Zhang and
Pfister~\cite{ZhangPfister_09} and by Lu \etal~\cite{Montanari} use density
evolution arguments to determine the precise threshold constant for sparse
measurement matrices, but these are for message-passing decoding algorithms
which are often not robust to noise and approximate sparsity.

In contrast to Corollary~\ref{cor:translation1} that presented a strong bound,
the following corollary presents a so-called weak bound (\confer~the
discussion in Section~\ref{sec:role:BSC:1}), but with a better threshold.

\mbox{}

\begin{corollary}
  \label{cor:daskalakis:etal:1}

  Let $\dv \in \Zpp$. Consider a random measurement matrix $\matrHCS \in \{ 0,
  1 \}^{m \times n}$ formed by placing $\dv$ random ones in each column, and
  zeros elsewhere. This measurement matrix succeeds in recovering a randomly
  supported $k = \alpha n$ sparse vector with probability $1 - o(1)$ if
  $\alpha$ is below some threshold value $\alpha_{m} (\dv, m/n)$.
\end{corollary}

\begin{IEEEproof}
  The result is obtained by combining
  Lemma~\ref{lemma:equation:nullspace:to:fc:1}
  with~\cite[Theorem~1]{DDKW07}. The latter paper also contains a way to
  compute the achievable threshold values $\alpha_{m} (\dv, m/n)$.
\end{IEEEproof}

\mbox{}

Using results about expander graphs from~\cite{DDKW07}, the above corollary
implies, for example, that for $m / n = 1 / 2$ and $\dv = 8$, a random
measurement matrix will recover with high probability a $k = \alpha n$ sparse
vector with random support if $\alpha \leq 0.002$. This is, of course, a much
higher threshold compared to the one presented above, but it only holds with
high probability over the vector support (therefore it is a so-called weak
bound). To the best of our knowledge, this is the first weak bound obtained
for random sparse measurement matrices under basis pursuit. 

The best thresholds known for LP decoding were recently obtained by Arora,
Daskalakis, and Steurer~\cite{ADS:Improved_LP} but require matrices that are
both left and right regular and also have logarithmically growing
girth.\footnote{However, as shown in~\cite{Vontobel:10:3}, these requirements
  on the left and right degrees can be significantly relaxed.} A random
bipartite matrix will not have logarithmically growing girth but there are
explicit deterministic constructions that achieve this (for example the
construction presented in Gallager's thesis~\cite[Appendix~C]{Gallager:63}).

\mbox{}

\begin{corollary} 
  \label{cor:arora:etal:1}

  Let $\dv, \dc \in \Zpp$. Consider a measurement matrix $\matrHCS \in \{ 0, 1
  \}^{m \times n}$ whose Tanner graph is a $(\dv, \dc)$-regular bipartite
  graph with $\Omega(\log n)$ girth. This measurement matrix succeeds in
  recovering a randomly supported $k = \alpha n$ sparse vector with
  probability $1 - o(1)$ if $\alpha$ is below some threshold function
  $\alpha'_{m}(\dv, \dc, m/n)$.
\end{corollary}

\begin{IEEEproof}
  The result is obtained by combining
  Lemma~\ref{lemma:equation:nullspace:to:fc:1} with~\cite[Theorem
  1]{ADS:Improved_LP}. The latter paper also contains a way to compute the
  achievable threshold values $\alpha'_{m} (\dv, \dc, m/n)$.
\end{IEEEproof}

\mbox{}

Using results from~\cite{ADS:Improved_LP}, the above corollary yields for $m /
n = 1 / 2$ and a $(3,6)$-regular Tanner graph with logarithmic girth (obtained
from Gallager's construction) the fact that sparse vectors with sparsity $k =
\alpha n$ are recoverable with high probability for $\alpha \leq
0.05$. Therefore, zero-one measurement matrices based on Gallager's
deterministic LDPC construction form sparse measurement matrices with an
order-optimal number of measurements (and the best known constants) for the
\CSLPD~/ basis pursuit setup.

\textbf{A note on deterministic constructions:} We say that a method to
construct a measurement matrix is deterministic if it can be created
deterministically in polynomial time, or it has a property that can be
verified in polynomial time. Unfortunately, all known bipartite
expansion-based constructions are non-deterministic because even though random
constructions will have the required expansion with high probability, there
is, to the best of our knowledge, no known efficient way to check expansion
above $\delta > 1/2$.  Similarly, there are no known ways to verify the
nullspace property or the restricted isometry property of a given candidate
measurement matrix in polynomial time.

There are several deterministic constructions of sparse measurement
matrices~\cite{Bajwa:Calderbank:Jafarpour:10:1, DeVore} which, however, would
require a slightly sub-optimal number of measurements (\ie, $m$ growing
super-linearly as a function of $n$ for $k = \alpha n$). The benefit of such
constructions is that reconstruction can be performed via algorithms that are
more efficient than generic convex optimization. To the best of our knowledge,
there are no previously known constructions of deterministic measurement
matrices with an optimal number of rows~\cite{Gilbert:Indyk:10:1}. The best
known constructions rely on explicit expander
constructions~\cite{CapalboExpanders, Guruswami:Umans:Vadhan:07:1}, but have
slightly sub-optimal parameters~\cite{Gilbert:Indyk:10:1, GIKS_RIP}. Our
construction of Corollary~\ref{cor:arora:etal:1} seems to be the first optimal
deterministic construction.

One important technical innovation that arises from the machinery we develop
is that \emph{girth} can be used to certify good measurement matrices. Since
checking and constructing high-girth graphs is much easier than constructing
graphs with high expansion, we can obtain very good deterministic measurement
matrices. For example, we can use Gallager's construction of LDPC matrices
with logarithmic girth to obtain sparse zero-one measurement matrices with an
order-optimal number of measurements under basis pursuit. The transition from
expansion-based arguments to girth-based arguments was achieved for the
channel coding problem in~\cite{Koetter:Vontobel:06:1}, then simplified and
brought to a new analytical level by Arora \etal~in~\cite{ADS:Improved_LP},
and afterwards generalized in~\cite{Vontobel:10:3}. Our connection results
extend the applicability of these results to compressed sensing.

We note that Corollary~\ref{cor:arora:etal:1} yields a weak bound, \ie, the
recovery of almost all $k$-sparse signals and therefore does not guarantee
recovering all $k$-sparse signals as the Capalbo \etal~\cite{CapalboExpanders}
construction (in conjunction with Corollary~\ref{cor:translation1}) would
ensure. On the other hand, girth-based constructions have constants that are
orders of magnitude higher than the ones obtained by random expanders. Since
the construction of~\cite{CapalboExpanders} gives constants that are worse
than the ones for random expanders, it seems that girth-based measurement
matrices have significantly higher provable thresholds of recovery. Finally,
we note that following~\cite{ADS:Improved_LP}, logarithmic girth $\Omega(\log
n )$ will yield a probability of failure decaying exponentially in the matrix
size $n$. However, even the much smaller girth requirement $\Omega (\log\log
n)$ is sufficient to make the probability of error decay as an inverse
polynomial of $n$.

A final remark: Chandar~\cite{Chandar:08:1} showed that zero-one measurement
matrices cannot have an optimal number of measurements if they must satisfy
the restricted isometry property for the $\ell_2$ norm. Note that this does
not contradict our work, since, as mentioned earlier on, RIP is just a
sufficient condition for signal recovery.

\subsection{Comments on Dense Measurement Matrices}
\label{sec:dense:measurement:matrices:1}

We conclude this section with some considerations about dense measurement
matrices, highlighting our current understanding that the translation of
positive performance guarantees from \CCLPD\ to \CSLPD\ displays the following
behavior: the denser a measurement matrix is, the weaker the translated
performance guarantees are.

\begin{remark}
  \label{remark:dense:measurement:matrix:1}

  Consider a randomly generated $m \times n$ measurement matrix $\matrHCS$
  where every entry is generated i.i.d.\ according to the distribution
  \begin{align*}
    \begin{cases}
      +1           & \text{with probability $1/6$} \\ 
      \phantom{+}0 & \text{with probability $2/3$} \\ 
      -1           & \text{with probability $1/6$}
    \end{cases}.
  \end{align*}
  This matrix, after multiplying it by the scalar $\sqrt{3/n}$, has the
  restricted isometry property (RIP) with high
  probability. (See~\cite{Baraniuk:Davenport:DeVore:Wakin:08:1}, which proves
  this property based on results in~\cite{Achlioptas:01:1}, which in turn
  proves that this family of matrices has a non-zero threshold.) On the other
  hand, one can show that the family of parity-check matrices where every
  entry is generated i.i.d.\ according to the distribution
  \begin{align*}
    \begin{cases}
      1 & \text{with probability $1/3$} \\ 
      0 & \text{with probability $2/3$}
    \end{cases}
  \end{align*}
  does \emph{not} have a non-zero threshold under \CCLPD\ for the
  BSC~\cite{Vontobel:Koetter:06:2}.
\end{remark}

Therefore, we conclude that the connection between \CSLPD\ and \CCLPD\ given
by Lemma~\ref{lemma:equation:nullspace:to:fc:1:complex:case} (an extension of
Lemma~\ref{lemma:equation:nullspace:to:fc:1} that is discussed in
Appendix~\ref{sec:extensions:bridge:lemma:1}) is not tight for dense matrices,
in the sense that the performance of \CSLPD\ for dense measurement matrices
can be much better than predicted by the translation of performance results
for \CCLPD\ of the corresponding parity-check matrix.

\section{Reformulations based on Graph Covers}
\label{sec:reformulations:1}

The aim of this section is to tighten the already close formal relationship
between \CCLPD\ and \CSLPD\ with the help of (topological) graph
covers~\cite{Massey:77:1, Stark:Terras:96:1}. We will see that the so-called
(blockwise) graph-cover decoder~\cite{Vontobel:Koetter:05:1:subm} (see
also~\cite{Vontobel:10:7:subm}), which is equivalent to \CCLPD\ and which can
be used to explain the close relationship between \CCLPD\ and message-passing
iterative decoding algorithms like the min-sum algorithm, can be translated to
the \CSLPD\ setup.

\begin{figure}
  \begin{center}
    \epsfig{file=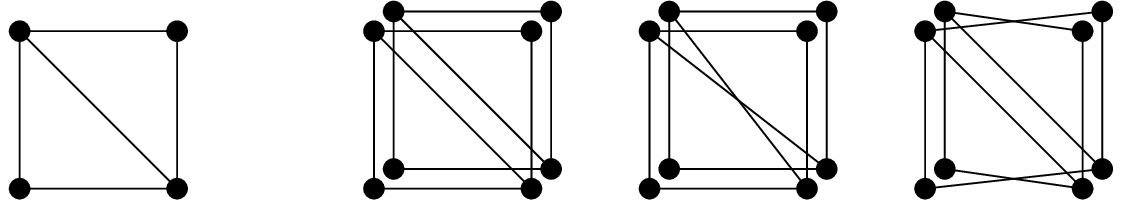, width=8cm} \\[1cm]
    \epsfig{file=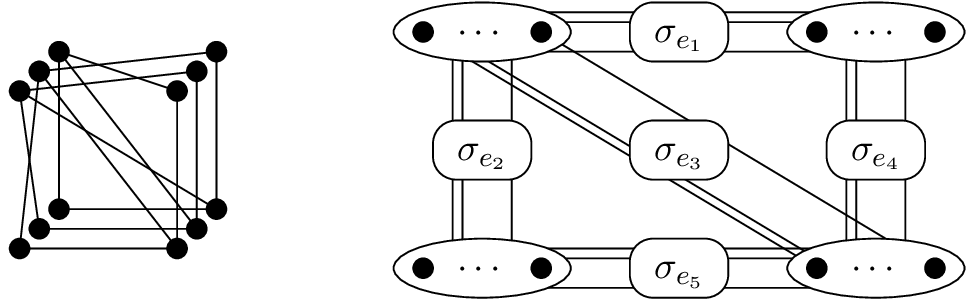, width=8cm}
  \end{center}
  \caption{Top left: base graph $\graph{G}$. Top right: a sample of possible
    $2$-covers of $\graph{G}$. Bottom left: a possible $3$-cover of
    $\graph{G}$. Bottom right: a possible $M$-cover of $\graph{G}$. Here,
    $\sigma_{e_1}, \ldots, \sigma_{e_5}$ are arbitrary edge permutations.}
  \label{fig:graph:cover:samples:1}
\end{figure}

\begin{figure}
  \begin{center}
    \epsfig{figure=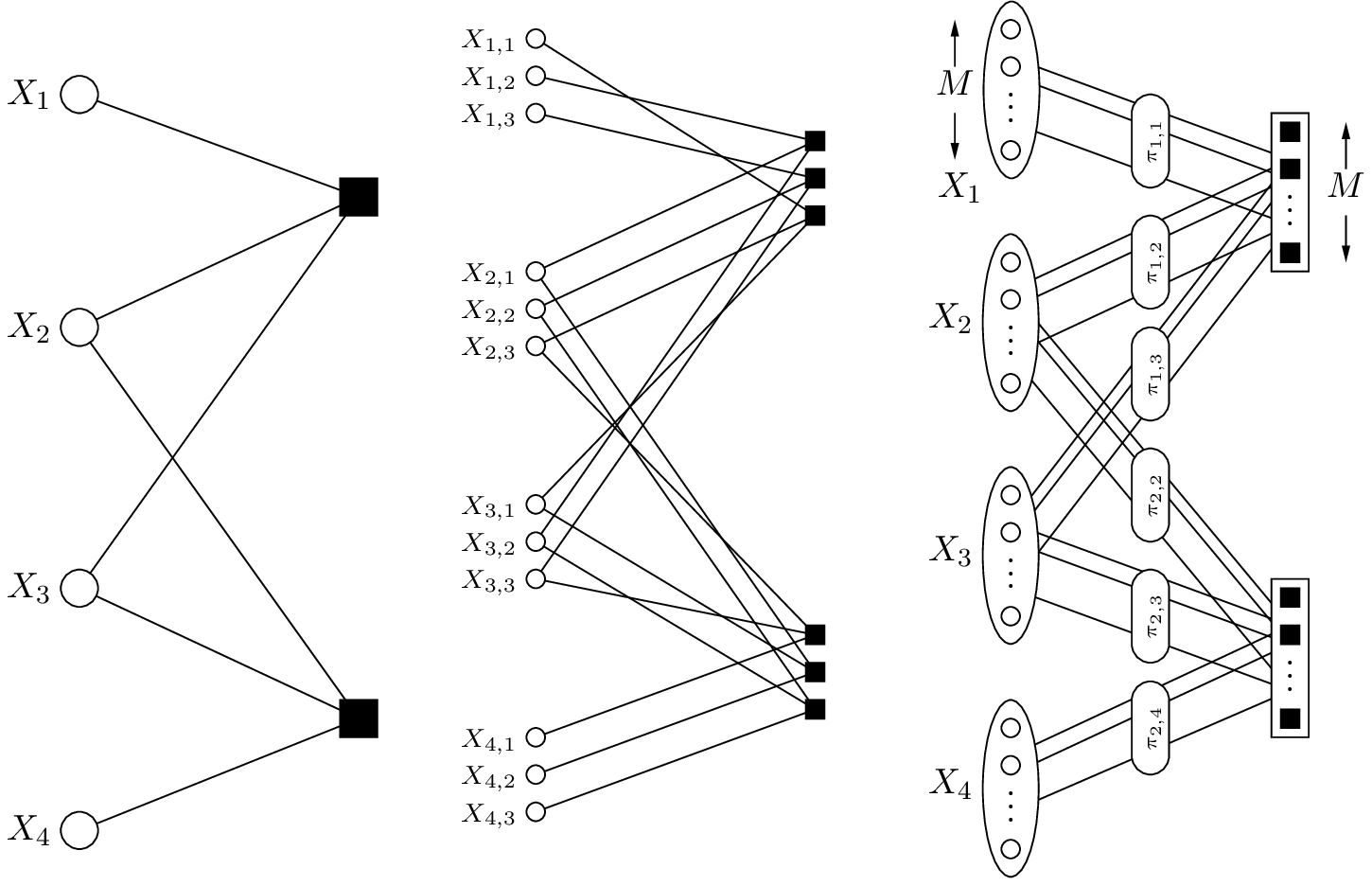,width=8.7cm}
  \end{center}
  \caption{Left: Tanner graph $\tgraph{T}{H}$. Middle: a possible $3$-cover of
    $\tgraph{T}{H}$. Right: a possible $M$-cover of $\tgraph{T}{H}$. Here, $\{
    \pi_{j,i} \}_{j,i}$ are arbitrary edge permutations.}
  \label{fig:simple:code:1}
\end{figure}

For an introduction to graph covers in general, and the graph-cover decoder in
particular, see~\cite{Vontobel:Koetter:05:1:subm}.
Figures~\ref{fig:graph:cover:samples:1} and~\ref{fig:simple:code:1} (taken
from~\cite{Vontobel:Koetter:05:1:subm}) show the main idea behind graph
covers. Namely, Figure~\ref{fig:graph:cover:samples:1} shows possible graph
covers of some (general) graph and Figure~\ref{fig:simple:code:1} shows
possible graph covers of some Tanner graph.

Note that in this section the compressed sensing setup will be over the
complex numbers. Also, the entries of the size-$m \times n$ measurement matrix
$\matrHCS$ will be allowed to take on any value in $\C$, \ie, the entries of
$\matrHCS$ are not restricted to have absolute value equal to zero or one.
Moreover, as in Section~\ref{sec:cc:lpd:1}, the channel coding problem assumes
an arbitrary binary-input output-symmetric memoryless channel, of which the
binary-input additive white Gaussian noise (AWGN) channel and the binary
symmetric channel (BSC) are prominent examples. As before, $\vx \in \{ 0, 1
\}^n$ will be the sent vector, $\vy \in \set{Y}^n$ will be the received
vector, and $\vlambda \in \R^n$ will contain the log-likelihood ratios
$\lambda_i \defeq \lambda_i(y_i) \defeq \log \bigl( \frac{P_{Y |
    X}(y_i|0)}{P_{Y | X}(y_i | 1)} \bigr)$, $i \in \setI(\matrHCS)$.

The rest of this section is organized as follows. In
Sections~\ref{sec:reformulations:of:CCMLD:1}
and~\ref{sec:reformulations:of:CCLPD:1} we show a variety of reformulations of
\CCMLD\ and \CCLPD, respectively. In particular, the latter subsection shows
reformulations of \CSLPD\ in terms of graph covers. Switching to compressed
sensing, in Section~\ref{sec:reformulations:of:CSOPT:1} we discuss
reformulations of \CSOPT\ that allow to see the close relationship of \CCMLD\
and \CSOPT. Afterwards, in Section~\ref{sec:reformulations:of:CSLPD:1}, we
present reformulations of \CSLPD\ which highlight the close connections, and
also the differences, between \CCLPD\ and \CSLPD.

\subsection{Reformulations of \CCMLD}
\label{sec:reformulations:of:CCMLD:1}

This subsection discusses several reformulations of \CCMLD, first for general
binary-input output-symmetric memoryless channels, then for the BSC. We start
by repeating two reformulations of \CCMLD\ from Section~\ref{sec:cc:lpd:1}.

\optprognoframe
{
\begin{alignat*}{2}
  \CCMLDver{1}:
  \quad
  &
  \text{minimize} \quad
  &&
    \langle \vlambda, \vx' \rangle \\
  &
  \text{subject to } \quad
  &&
    \vx' \in \codeCCC. \\[0.5cm]
  \CCMLDver{2}:
  \quad
  &
  \text{minimize} \quad
  &&
    \langle \vlambda, \vx' \rangle \\
  &
  \text{subject to } \quad
  &&
    \vx' \in \convhull(\codeCCC).
\end{alignat*}
}{-0.25cm}

Towards yet another reformulation of \CCMLD\ that we would like to present in
this subsection, it is useful to introduce the hard-decision vector $\ovy$,
along with the syndrome vector $\vs$ induced by $\ovy$.

\begin{definition}
  \label{def:syndrome:for:general:binary:input:channels:1}

  Let $\ovy \in \GF{2}^n$ be the hard-decision vector based on the
  log-likelihood ratio vector $\vlambda$, namely let
  \begin{align*}
    \oy_i
      &\defeq
         \begin{cases}
           0 & \text{if $\lambda_i > 0$} \\
           1 & \text{if $\lambda_i < 0$} \\
         \end{cases}
         \quad \text{(for all $i \in \setI$).}
  \end{align*}
  (If $\lambda_i = 0$, we set $\oy_i \defeq 0$ or $\oy_i \defeq 1$ according
  to some deterministic or random rule.) Moreover, let
  \begin{align*}
    \vs
      &\defeq
         \matrHCC \cdot \ovy \ \inGFtwo
  \end{align*}
  be the syndrome induced by $\ovy$.
\end{definition}

Clearly, if the channel under consideration is a BSC with cross-over
probability smaller than $1/2$ then $\ovy = \vy$.

With this, we have for any binary-input output-symmetric memoryless channel
the following reformulation of \CCMLD\ in terms of $\ve' \defeq \vy - \vx' \
\inGFtwo$.

\optprog
{
\begin{alignat*}{2}
  \CCMLDver{3}:
  \quad
  &
  \text{minimize} \quad
  &&
    \onenorm{\vlambda_{\supp(\ve')}} \\
  &
  \text{subject to } \quad
  &&
    \matrHCC \cdot \ve' = \vs \ \inGFtwo.
\end{alignat*}
}{-0.8cm}

Clearly, once the error vector estimate $\ve'$ is found, the codeword estimate
$\vx'$ is obtained with the help of the expression $\vx' = \vy - \ve' \
\inGFtwo$.

Note that for the special case of a binary-input AWGNC, this reformulation can
be found, for example, in~\cite{Breitbach:Bossert:Lucas:Kempter:98:1}
or~\cite[Chapter~10]{Lin:Costello:04:1}.

\mbox{}

\begin{theorem}
  \label{theorem:mld:reformulation:1}

  \CCMLDver{3} is a reformulation of \CCMLDver{1}.
\end{theorem}

\begin{IEEEproof}
  See Appendix~\ref{sec:proof:theorem:mld:reformulation:1}.
\end{IEEEproof}

\mbox{}

For a BSC we can specialize the above reformulations. Namely, for a BSC with
cross-over probability $\varepsilon$, $0 \leq \varepsilon < 1/2$, we have
$\absnorm{\lambda_i} = L$, $i \in \setI$, where $L \defeq \log\bigl(
\frac{1-\varepsilon}{\varepsilon} \bigr) > 0$. Then, with a slight abuse of
notation by employing $\onenorm{\,\cdot\,}$ also for vectors over $\GF{2}$, we
obtain the following reformulation.

\optprognoframe
{
\begin{alignat*}{2}
  \CCMLDver{4}\ \text{(BSC)}:
  \quad
  &
  \text{minimize} \quad
  &&
    \onenorm{\ve'} \\
  &
  \text{subject to } \quad
  &&
    \matrHCC \cdot \ve' = \vs \ \inGFtwo.
\end{alignat*}
}{-0.25cm}

\noindent
Moreover, with a slight abuse of notation by employing $\zeronorm{\,\cdot\,}$
also for vectors over $\GF{2}$, $\CCMLDver{4}\ \text{(BSC)}$ can be written as
follows.

\optprognoframe
{
\begin{alignat*}{2}
  \CCMLDver{5}\ \text{(BSC)}:
  \quad
  &
  \text{minimize} \quad
  &&
    \zeronorm{\ve'} \\
  &
  \text{subject to } \quad
  &&
    \matrHCC \cdot \ve' = \vs \ \inGFtwo.
\end{alignat*}
}{-0.8cm}

\subsection{Reformulations of \CCLPD}
\label{sec:reformulations:of:CCLPD:1}

We start by repeating the definition of \CCLPD\ from
Section~\ref{sec:cc:lpd:1}.

\optprognoframe
{
\begin{alignat*}{2}
  \CCLPD:
  \quad
  &
  \text{minimize} \quad
  &&
    \langle \vlambda, \vx' \rangle \\
  &
  \text{subject to } \quad
  &&
    \vx' \in \fp{P}(\matrHCC).
\end{alignat*}
}{-0.25cm}

\noindent
The aim of this subsection is to discuss various reformulations of \CCLPD\ in
terms of graph covers. In particular, the following reformulation of \CCLPD\
was presented in~\cite{Vontobel:Koetter:05:1:subm} and was called (blockwise)
graph-cover decoding.

\optprognoframe
{
\begin{alignat*}{2}
  \CCLPDver{1}:
  \quad
  &
  \text{minimize} \quad
  &&
    \frac{1}{M} \cdot \left\langle \vlambda^{\uparrow M}, \tvx' \right\rangle \\
  &
  \text{subject to } \quad
  &&
    \tmatrHCC \cdot \tvx' = \vzero^{\uparrow M} \ \inGFtwo.
\end{alignat*}
}{-0.25cm}

\noindent Here the minimization is over all $M \in \Zpp$ and over all
parity-check matrices $\tmatrHCC$ induced by all possible $M$-covers of the
Tanner graph of $\matrHCC$.\footnote{Note that here $\tmatrHCC$ is obtained by
  the standard procedure to construct a graph
  cover~\cite{Vontobel:Koetter:05:1:subm}, and not by the procedure in
  Definition~\ref{def:measurement:matrix:cover:1}
  (\confer~Appendix~\ref{sec:extensions:bridge:lemma:1}).}

Using the same line of reasoning as in
Section~\ref{sec:reformulations:of:CCMLD:1}, \CCLPD\ can be rewritten as
follows.

\optprog
{
\begin{alignat*}{2}
  \CCLPDver{2}:
  \quad
  &
  \text{minimize} \quad
  &&
    \frac{1}{M} \cdot \onenorm{\vlambda^{\uparrow M}_{\supp(\tve')}} \\
  &
  \text{subject to } \quad
  &&
    \tmatrHCC \cdot \tve' = \vs^{\uparrow M} \ \inGFtwo.
\end{alignat*}
}{-0.8cm}

\noindent Again, the minimization is over all $M \in \Zpp$ and over all
parity-check matrices $\tmatrHCC$ induced by all possible $M$-covers of the
Tanner graph of $\matrHCC$.

For the BSC with cross-over probability $\varepsilon$, $0 \leq \varepsilon <
1/2$, we get, with a slight abuse of notation as in
Section~\ref{sec:reformulations:of:CCMLD:1}, the following specialized
results.

\optprognoframe
{
\begin{alignat*}{2}
  \CCLPDver{3}\ \text{(BSC)}:
  \
  &
  \text{minimize} \quad
  &&
    \frac{1}{M} \cdot \onenorm{\tve'} \\
  &
  \text{subject to } \quad
  &&
    \tmatrHCC \cdot \tve' = \vs^{\uparrow M} \ \inGFtwo.
\end{alignat*}
}{-0.25cm}

\optprognoframe
{
\begin{alignat*}{2}
  \CCLPDver{4}\ \text{(BSC)}:
  \
  &
  \text{minimize} \quad
  &&
    \frac{1}{M} \cdot \zeronorm{\tve'} \\
  &
  \text{subject to } \quad
  &&
    \tmatrHCC \cdot \tve' = \vs^{\uparrow M} \ \inGFtwo.
\end{alignat*}
}{-0.25cm}

\subsection{Reformulations of \CSOPT}
\label{sec:reformulations:of:CSOPT:1}

We start by repeating the definition of \CSOPT\ from
Section~\ref{sec:cs:lpd:1}.

\optprognoframe
{
\begin{alignat*}{2}
  \CSOPT:
  \quad
  &
  \text{minimize} \quad
  &&
    \zeronorm{\ve'} \\
  &
  \text{subject to } \quad
  &&
    \matrHCS \cdot \ve' = \vs.
\end{alignat*}
}{-0.25cm}

\noindent Clearly, this is formally very similar to \CCMLDver{5}\ (BSC).

In order to show the tight formal relationship of \CSOPT\ with \CCMLD\ for
general binary-input output-symmetric memoryless channels, in particular with
respect to the reformulation \CCMLDver{3}, we rewrite \CSOPT\ as follows.

\optprog
{
\begin{alignat*}{2}
  \CSOPTver{1}:
  \quad
  &
  \text{minimize} \quad
  &&
    \onenorm{\vone_{\supp(\ve')}} \\
  &
  \text{subject to } \quad
  &&
   \matrHCS \cdot \ve' = \vs.
\end{alignat*}
}{-0.8cm}

\subsection{Reformulations of \CSLPD}
\label{sec:reformulations:of:CSLPD:1}

We now come to the main part of this section, namely the reformulation of
\CSLPD\ in terms of graph covers. We start by repeating the definition of
\CSLPD\ from Section~\ref{sec:cs:lpd:1}.

\optprognoframe
{
\begin{alignat*}{2}
  \CSLPD:
  \quad
  &
  \text{minimize} \quad
  &&
    \onenorm{\ve'} \\
  &
  \text{subject to } \quad
  &&
    \matrHCS \cdot \ve' = \vs.
\end{alignat*}
}{-0.25cm}

\noindent
As shown in the upcoming Theorem~\ref{theorem:CSLPD:reformulation:2:vs:1},
\CSLPD\ can be rewritten as follows.

\optprog
{
\begin{alignat*}{2}
  \CSLPDver{1}:
  \quad
  &
  \text{minimize} \quad
  &&
    \frac{1}{M} \cdot \onenorm{\tve'} \\
  &
  \text{subject to } \quad
  &&
    \tmatrHCS \cdot \tve' = \vs^{\uparrow M}.
\end{alignat*}
}{-0.8cm}

\noindent
Here the minimization is over all $M \in \Zpp$ and over all measurement
matrices $\tmatrHCS$ induced by all possible $M$-covers of the Tanner graph of
$\matrHCS$.

\mbox{}

\begin{theorem}
  \label{theorem:CSLPD:reformulation:2:vs:1}

  \CSLPDver{1} is a reformulation of \CSLPD.
\end{theorem}

\begin{IEEEproof}
  See Appendix~\ref{sec:proof:theorem:CSLPD:reformulation:2:vs:1}.
\end{IEEEproof}

\mbox{}

Clearly, \CSLPDver{1} is formally very close to \CCLPDver{3} (BSC), thereby
showing that graph covers can be used to exhibit yet another tight formal
relationship between \CSLPD\ and \CCLPD.

Nevertheless, these graph-cover based reformulations also highlight
differences between the relaxation used in the context of channel coding and
the relaxation used in the context of compressed sensing.

\begin{itemize}

\item When relaxing \CCMLD\ to obtain \CCLPD, the cost function remains the
  same (call this property $\mathrm{P}1$) but the domain is relaxed (call this
  property $\mathrm{P}2$). In the graph-cover reformulations of \CCLPD,
  property $\mathrm{P}1$ is reflected by the fact that the cost function is a
  straightforward generalization of the cost function for \CCMLD. Property
  $\mathrm{P}2$ is reflected by the fact that in general there are feasible
  vectors in graph covers that cannot be explained as liftings of (convex
  combinations of) feasible vectors in the base graph and that, for suitable
  $\vlambda$-vectors, have strictly lower cost function values than any
  feasible vector in the base graph.

\item When relaxing \CSOPT\ to obtain \CSLPD, the cost function is changed
  (call this property $\mathrm{P}1'$), but the domain remains the same (call
  this property $\mathrm{P}2'$). In the graph-cover reformulations of \CSLPD,
  property $\mathrm{P}1'$ is reflected by the fact that the cost function is
  \emph{not} a straightforward generalization of the cost function of
  \CSOPT. Property $\mathrm{P}2'$ is reflected by the fact that feasible
  vectors in graph covers are such that they \emph{do not} yield cost function
  values that are smaller than the cost function value of the best feasible
  vector in the base graph.

\end{itemize}

\section{Minimizing the Zero-Infinity Operator}
\label{sec:minimizing:zero:infty:norm:1}

For any real vector $\va$ we define the zero-infinity operator to be
\begin{align*}
  \zeroinfoperator{\va}
    &\defeq
       \zeronorm{\va}
       \cdot
       \infnorm{\va},
\end{align*}
\ie, the product of the zero norm $\zeronorm{\va} = \card{\supp(\va)}$ of
$\va$ and of the infinity norm $\infnorm{\va} = \max_i |a_i|$ of $\va$. Note
that for any $c \in \C$ and any real vector $\va$ it holds that
$\zeroinfoperator{c \cdot \va} = \absnorm{c} \cdot \zeroinfoperator{\va}$.

Based on this operator, in the present section we introduce $\CSOPTzeroinf$,
and we show, with the help of graph covers, that \CSLPD\ can not only be seen
as a relaxation of \CSOPT\ but also as a relaxation of $\CSOPTzeroinf$. We do
this by proposing a relaxation of $\CSOPTzeroinf$, called $\CSRELzeroinf$, and
by then showing that $\CSRELzeroinf$ is equivalent to \CSLPD.

Moreover, we argue that the solution of \CSLPD\ is ``closer'' to the solution
of $\CSOPTzeroinf$ than the solution of \CSLPD\ is to the solution of
\CSOPT. Note that similar to \CSOPT, the problem $\CSOPTzeroinf$ is in general
an intractable optimization problem.

One motivation for looking for different problems whose relaxations equals
\CSLPD\ is to better understand the ``strengths'' and ``weaknesses'' of
\CSLPD. In particular, if \CSLPD\ is the relaxation of two different problems
(like \CSOPT\ and $\CSOPTzeroinf$), but these two problems yield different
solutions, then the solution of the relaxed problem will disagree with the
solution of at least one of the two problems.

This section is structured as follows. We start by defining $\CSOPTzeroinf$ in
Section~\ref{sec:def:CSOPTzeroinf:1}. Then, in
Section~\ref{sec:geometrical:aspects:of:CSOPTzeroinf:1}, we discuss some
geometrical aspects of $\CSOPTzeroinf$, in particular with respect to the
geometry behind \CSOPT\ and \CSLPD. Finally, in
Section~\ref{sec:relaxation:of:CSOPTzeroinf:1}, we introduce $\CSRELzeroinf$
and show its equivalence to \CSLPD.

\subsection{Definition of $\CSOPTzeroinf$}
\label{sec:def:CSOPTzeroinf:1}

The optimization problem $\CSOPTzeroinf$ is defined as follows.

\optprog
{
\begin{alignat*}{2}
  \CSOPTzeroinf:
  \quad
  &
  \text{minimize} \quad
  &&
    \zeroinfoperator{\ve'} \\
  &
  \text{subject to } \quad
  &&
    \matrHCS \cdot \ve' = \vs.
\end{alignat*}
}{-0.8cm}

\noindent
Whereas the cost function of \CSOPT, \ie, $\zeronorm{\ve'}$, measures the
sparsity of $\ve'$ but not the magnitude of the elements of $\ve'$, the cost
function of $\CSOPTzeroinf$, \ie, $\zeroinfoperator{\ve'}$, represents a
trade-off between measuring the sparsity of $\ve'$ and measuring the largest
magnitude of the components of $\ve'$. Clearly, in the same way that there are
many good reasons to look for the vector $\ve'$ that minimizes the zero-norm
(among all $\ve'$ that satisfy $\matrHCS \cdot \ve' = \vs$), there are also
many good reasons to look for the vector $\ve'$ that minimizes the
zero-infinity operator (among all $\ve'$ that satisfy $\matrHCS \cdot \ve' =
\vs$). In particular, the latter is attractive when we are looking for a
sparse vector $\ve'$ that does not have an imbalance in magnitudes between the
largest component and the set of most important components.

With a slight abuse of notation, we can apply the zero-infinity operator
$\zeroinfoperator{\,\cdot\,}$ also to vectors over $\GF{2}$ and obtain the
following reformulation of \CCMLD\ (BSC). (Note that for any vector $\va$ over
$\GF{2}$ it holds that $\zeroinfoperator{\va} = \onenorm{\va} = \wH(\va)$.)

\optprognoframe
{
\begin{alignat*}{2}
  \CCMLDver{6}\ \text{(BSC)}:
  \quad
  &
  \text{minimize} \quad
  &&
    \zeroinfoperator{\ve'} \\
  &
  \text{subject to } \quad
  &&
    \matrHCC \cdot \ve' = \vs.
\end{alignat*}
}{-0.25cm}

\noindent
This clearly shows that there is a close formal relationship not only between
\CCMLD\ (BSC) and \CSOPT, but also between \CCMLD\ (BSC) and $\CSOPTzeroinf$.

\subsection{Geometrical Aspects of $\CSOPTzeroinf$}
\label{sec:geometrical:aspects:of:CSOPTzeroinf:1}

\begin{figure}
  \begin{center}
    \epsfig{file=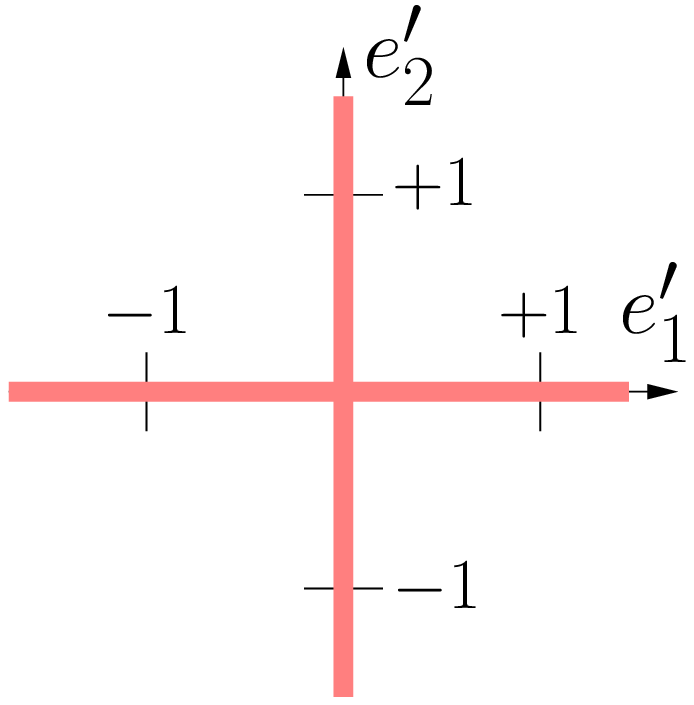,
            width=0.30\linewidth}
    \
    \epsfig{file=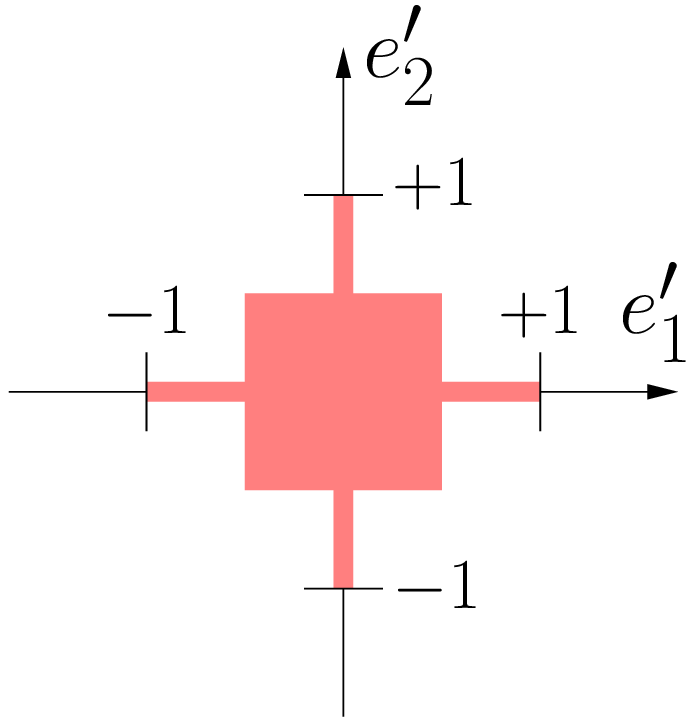,
            width=0.30\linewidth}
    \
    \epsfig{file=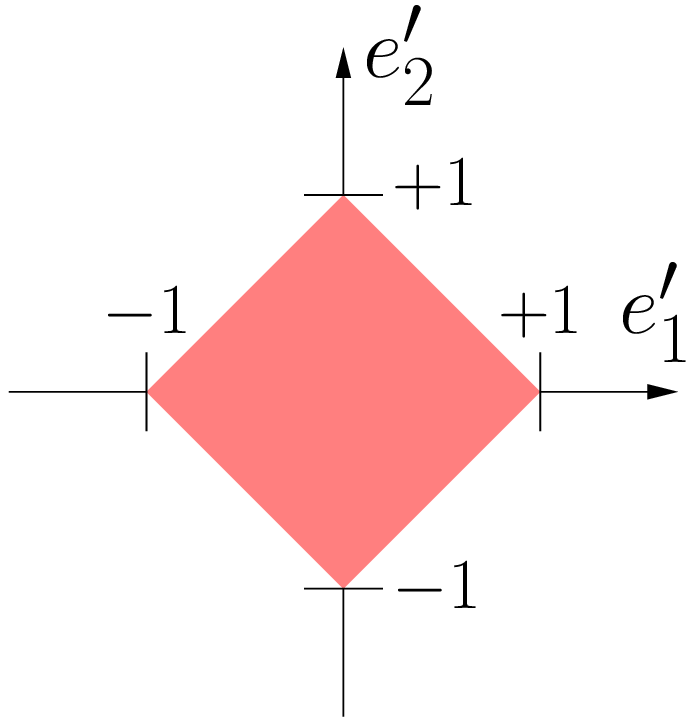,
            width=0.30\linewidth}
  \end{center}
  \caption{Unit balls for some operators. Left: $\bigl\{ \ve' \in \R^2 \bigm|
    \zeronorm{\ve'} \leq 1 \bigr\}$. Middle: $\bigl\{ \ve' \in \R^2 \bigm|
    \zeroinfoperator{\ve'} \leq 1 \bigr\}$. Right: $\bigl\{ \ve' \in \R^2 \bigm|
    \onenorm{\ve'} \leq 1 \bigr\}$.}
  \label{fig:unit:balls:1}
\end{figure}

We want to discuss some geometrical aspects of \CSOPT, $\CSOPTzeroinf$, and
\CSLPD. Namely, as is well known, \CSOPT\ can be formulated as finding the
smallest $\ell_0$-norm ball of radius $r$
(\confer~Figure~\ref{fig:unit:balls:1}~(left)) that intersects the set
$\bigl\{ \ve' \bigm| \matrHCS \cdot \ve' = \vs \bigr\}$, and in the same
spirit, \CSLPD\ can be formulated as finding the smallest $\ell_1$-norm ball
of radius $r$ (\confer~Figure~\ref{fig:unit:balls:1}~(right)) that intersects
with the set $\bigl\{ \ve' \bigm| \matrHCS \cdot \ve' = \vs \bigr\}$. Clearly,
the fact that \CSOPT\ and \CSLPD\ can yield different solutions stems from the
fact that these balls have different shapes. Of course, the success of \CSLPD\
is a consequence of the fact that, nevertheless, under suitable conditions,
the solution given by the $\ell_1$-norm ball is (nearly) the same as the
solution given by the $\ell_0$-norm ball.

In the same vein, $\CSOPTzeroinf$ can be formulated as finding the smallest
zero-infinity-operator ball of radius $r$
(\confer~Figure~\ref{fig:unit:balls:1}~(middle)) that intersects the set
$\bigl\{ \ve' \bigm| \matrHCS \cdot \ve' = \vs \bigr\}$. As it can be seen
from Figure~\ref{fig:unit:balls:1}, the zero-infinity-operator unit ball is
closer in shape to the $\ell_1$-norm unit ball than the $\ell_0$-norm unit
ball is to the $\ell_1$-norm unit ball. Therefore, we expect that the solution
given by \CSLPD\ is ``closer'' to the solution given by $\CSOPTzeroinf$ than
the solution of \CSLPD\ is to the solution given by \CSOPT. In that sense,
$\CSOPTzeroinf$ is at least as justifiably as \CSOPT\ a difficult optimization
problem whose solution is approximated by \CSLPD.

\subsection{Relaxation of $\CSOPTzeroinf$}
\label{sec:relaxation:of:CSOPTzeroinf:1}

In this subsection we introduce $\CSRELzeroinf$ as a relaxation of
$\CSOPTzeroinf$; the main result will be that $\CSRELzeroinf$ equals
\CSLPD. Our results will be formulated in terms of graph covers, we therefore
use the graph-cover related notation that was introduced in
Section~\ref{sec:reformulations:1}, along with the mapping $\varphiM$ that was
defined in Section~\ref{sec:notation:1}.

In order to motivate the formulation of $\CSRELzeroinf$, we first present a
reformulation of \CCLPDver\ (BSC). Namely, \CCLPDver{3}\ (BSC) or
\CCLPDver{4}\ (BSC) from Section~\ref{sec:reformulations:of:CCLPD:1} can be
rewritten as follows.

\optprognoframe
{
\begin{alignat*}{2}
  \CCLPDver{5}\ \text{(BSC)}:
  \
  &
  \text{minimize} \quad
  &&
    \frac{1}{M} \cdot \zeroinfoperator{\tve'} \\
  &
  \text{subject to } \quad
  &&
    \tmatrHCC \cdot \tve' = \vs^{\uparrow M} \ \inGFtwo.
\end{alignat*}
}{-0.25cm}

\noindent
Then, because for any vector $\tvs \in \GF{2}^{|\setJ| \cdot M}$ it holds that
$\varphiM(\tvs) = \vs$ if and only if $\tvs = \vs^{\uparrow M}$, \CCLPDver{5}
(BSC) can also be written as follows.

\optprognoframe
{
\begin{alignat*}{2}
  \CCLPDver{6}\ \text{(BSC)}:
  \quad
  &
  \text{minimize} \quad
  &&
    \frac{1}{M} \cdot \zeroinfoperator{\tve'} \\
  &
  \text{subject to } \quad
  &&
    \tmatrHCC \cdot \tve' = \tvs \ \inGFtwo \\
  &&&
    \hskip+0.275cm \varphiM(\tvs) = \vs.
\end{alignat*}
}{-0.8cm}

\noindent
The transition that leads from \CCMLD\ to its relaxation \CCLPDver{6}\ (BSC)
inspires a relaxation of $\CSOPTzeroinf$ as follows.

\optprog
{
\begin{alignat*}{2}
  \CSRELzeroinf:
  \quad
  &
  \text{minimize} \quad
  &&
    \frac{1}{M} \cdot \zeroinfoperator{\tve'} \\
  &
  \text{subject to } \quad
  &&
    \tmatrHCS \cdot \tve' = \tvs \\
  &&&
    \hskip+0.24cm \varphiM(\tvs) = \vs.
\end{alignat*}
}{-0.8cm}

\noindent
Here the minimization is over all $M \in \Zpp$ and over all measurement
matrices $\tmatrHCS$ induced by all possible $M$-covers of the Tanner graph of
$\matrHCS$. Note that, in contrast to \CCLPDver{6}\ (BSC), in general the
optimal solution $(\tve,\tvs)$ of $\CSRELzeroinf$ does \emph{not} satisfy
$\tvs = \vs^{\uparrow M}$.

Towards establishing the equivalence of $\CSRELzeroinf$ and \CSLPD, the
following simple lemma will prove to be useful.

\mbox{}

\begin{lemma}
  \label{lemma:onenorm:zeroinfoperator:relationship:1}

  For any real vector $\va$ it holds that
  \begin{align*}
    \onenorm{\va}
      &\leq \zeroinfoperator{\va},
  \end{align*}
  with equality if and only if all non-zero components of $\va$ have the same
  absolute value.
\end{lemma}

\begin{IEEEproof}
  The proof of this lemma is straightforward.
\end{IEEEproof}

\mbox{}

\begin{theorem}
  \label{theorem:CSLPD:and:CSOPTzeroinfty:connection:1}

  Let $\matrHCS$ be a measurement matrix over the reals with entries equal to
  zero, one, and minus one. For syndrome vectors $\vs$ that have only rational
  components, \CSLPD\ and $\CSRELzeroinf$ are equivalent in the sense that
  there is an optimal $\ve'$ in \CSLPD\ and an optimal $\tve'$ in
  $\CSRELzeroinf$ such that $\ve' = \varphiM(\tve')$.
\end{theorem}

\begin{IEEEproof}
  See Appendix~\ref{sec:proof:theorem:CSLPD:and:CSOPTzeroinfty:connection:1}.
\end{IEEEproof}

\section{Conclusions and Outlook}
\label{sec:conclusions:1}

In this paper we have established a mathematical connection between channel
coding and compressed sensing LP relaxations. The key observation, in its
simplest version, was that points in the nullspace of a zero-one matrix
(considered over the reals) can be mapped to points in the fundamental cone of
the same matrix (considered as the parity-check matrix of a code over
$\GF{2}$). This allowed us to show, among other results, that parity-check
matrices of ``good'' channel codes can be used as provably ``good''
measurement matrices under basis pursuit.

Let us comment on a variety of topics.
\begin{itemize}

\item In addition to \CSLPD, a number of combinatorial algorithms
  (\eg~\cite{DM_SP09, XH_expander, Jafarpour:Xu:Hassibi:Calderbank:09:1,
    Tropp:Gilbert:07:1, GIKS_RIP, GLW_random}) have been proposed for
  compressed sensing problems, with the benefit of faster decoding complexity
  and comparable performance to \CSLPD.  It would be interesting to
  investigate if the connection of sparse recovery problems to channel coding
  extends in a similar manner for these decoders. One example of such a clear
  connection is the bit-flipping algorithm of Sipser and
  Spielman~\cite{Sipser:Spielman:96} and the corresponding algorithm for
  compressed sensing by Xu and
  Hassibi~\cite{XH_expander}. Channel-coding-inspired message-passing decoders
  for compressed sensing problems were also recently discussed
  in~\cite{Akcakaya:Tarokh:08:1, ZhangPfister_09, Montanari,
    Calderbank:Howard:Jafarpour:10:1, Babadi:Tarokh:11:1}.

\item An interesting research direction is to use optimized LDPC matrices
  (see, \eg~\cite{Richardson:Urbanke:08:1}) to create measurement
  matrices. There is a large body of channel coding work that could be
  transferable to the measurement matrix design problem.

  In this context, an important theoretical question is related to being able
  to certify in polynomial time that a given measurement matrix has ``good''
  performance. To the best of our knowledge, our results form the first known
  case where girth, an efficiently checkable property, can be used as a
  certificate of goodness of a measurement matrix. It is possible that girth
  can be used to establish a success witness for \CSLPD\ directly, and this
  would be an interesting direction for future research.

\item One important research direction in compressed sensing involves dealing
  with noisy measurements. This problem can still be addressed with $\ell_1$
  minimization (see, \eg, \cite{Candes:Plan:09}) and also with less complex
  signal reconstruction algorithms (see, \eg,
  \cite{Fletcher:Rangan:Goyal:09:1}). It would be very interesting to
  investigate if our nullspace connections can be extended to a coding theory
  result equivalent to noisy compressed sensing.

\item Beyond channel coding problems, the LP relaxation
  of~\cite{Feldman:Wainwright:Karger:05:1} is a special case of a relaxation
  of the marginal polytope for general graphical models. One very interesting
  research direction is to explore if the connection we have established
  between \CSLPD\ and \CCLPD\ is also just a special case of a more general
  theory.

\item We have also discussed various reformulations of the optimization
  problems under investigation. This leads to a strengthening of the ties
  between some of the optimization problems. Moreover, we have introduced the
  zero-infinity operator optimization problem $\CSOPTzeroinf$, an optimization
  problem with the property that the solution of \CSLPD\ can be considered to
  be at least as good an approximation of the solution of $\CSOPTzeroinf$ as
  the solution of \CSLPD\ is an approximation of the solution of \CSOPT. We
  leave it as an open question if the results and observations of
  Section~\ref{sec:minimizing:zero:infty:norm:1} can be generalized for more
  general matrices or specific families of signals (like non-negative sparse
  signals as in~\cite{Khajehnejad:Dimakis:Xu:Hassibi:11:1,
    Raginsky:Willett:Harmany:Marcia:10:1}).

\end{itemize}

\section*{Acknowledgments}

We would like to thank Babak Hassibi and Waheed Bajwa for stimulating
discussions with respect to the topic of this paper. Moreover, we greatly
appreciate the reviewers' comments that lead to an improved presentation of
the results.

\appendices

\section{Proof of
               Theorem~\ref{theorem:l1:l1:approximation:guarantee:1}}
\label{sec:proof:theorem:l1:l1:approximation:guarantee:1}

Suppose that $\matrHCS$ has the claimed nullspace property.
Since $\matrHCS \cdot \ve = \vs$ and $\matrHCS \cdot \hve = \vs$, it easily
follows that $\vnnu \defeq \ve - \hve$ is in the nullspace of
$\matrHCS$. So,
\begin{align}
  \onenorm{\veS}
  +
  \onenorm{\veSc}
    &= \onenorm{\ve} \nonumber \\
    &\overset{\text{(a)}}{\geq}
       \onenorm{\hve} \nonumber \\
    &= \onenorm{\ve - \vnnu} \nonumber \\
    &= \onenorm{\veS - \vnnuS}
       + 
       \onenorm{\veSc - \vnnuSc} \nonumber \\
    &\overset{\text{(b)}}{\geq}
       \onenorm{\veS}
       -
       \onenorm{\vnnuS}
       +
       \onenorm{\vnnuSc}
       -
       \onenorm{\veSc} \nonumber \\
    &\overset{\text{(c)}}{\geq}
       \onenorm{\veS}
       +
       \frac{C-1}{C+1}
         \cdot
         \onenorm{\vnnu}
       -
       \onenorm{\veSc},
         \label{eq:ell1:ell1:guarantee:1}
\end{align}
where step~$\text{(a)}$ follows from the fact that the solution of \CSLPD\
satisfies $\onenorm{\hve} \leq \onenorm{\ve}$, where step~$\text{(b)}$ follows
from applying the triangle inequality property of the $\ell_1$-norm twice, and
where step~$\text{(c)}$ follows from
\begin{align*}
  -
  \onenorm{\vnnuS}
  +
  \onenorm{\vnnuSc}
    &\overset{\text{(d)}}{\geq}
       \frac{C-1}{C+1}
       \cdot
       \onenorm{\vnnu}.
\end{align*}
Here, step~$\text{(d)}$ is a consequence of
\begin{align*}
  (C \! + \! 1)
    \, \cdot \, &
    \big(
      -
      \onenorm{\vnnuS}
      +
      \onenorm{\vnnuSc}
    \big) \\
    &= -
       C
         \cdot
         \onenorm{\vnnuS}
       -
       \onenorm{\vnnuS}
       +
       C
         \cdot
         \onenorm{\vnnuSc}
       +
       \onenorm{\vnnuSc} \\
    &\overset{\text{(e)}}{\geq}
       -
       \onenorm{\vnnuSc}
       -
       \onenorm{\vnnuS}
       +
       C
         \cdot
         \onenorm{\vnnuSc}
       +
       C
         \cdot
         \onenorm{\vnnuS} \\
    &= (C \! - \! 1)
         \cdot
         \onenorm{\vnnuS}
       +
       (C \! - \! 1)
         \cdot
         \onenorm{\vnnuSc} \\
    &= (C \! - \! 1)
         \cdot
         \onenorm{\vnnu},
\end{align*}
where step~$\text{(e)}$ follows from applying twice the fact that $\vnnu \in
\nullspaceR(\matrHCS)$ and the assumption that $\matrHCS \in
\NSPR(k,C)$. Subtracting the term $\onenorm{\veS}$ on both sides
of~\eqref{eq:ell1:ell1:guarantee:1}, and solving for $\onenorm{\vnnu} =
\onenorm{\ve - \hve}$ yields the promised result.

\section{Proof of
               Lemma~\ref{lemma:bsc:strict:balancedness:1}}
\label{sec:proof:lemma:bsc:strict:balancedness:1}

Without loss of generality, we can assume that the all-zero codeword was
transmitted. Let $+L > 0$ be the log-likelihood ratio associated with a
received $0$, and let $-L < 0$ be the log-likelihood ratio associated with a
received $1$. Therefore, $\lambda_i = +L$ if $i \in \setSc$ and $\lambda_i =
-L$ if $i \in \setS$. Then it follows from the assumptions in the lemma
statement that for any $\vomega \in \fc{K}(\matrHCC) \setminus \{ \vzero \}$
it holds that
\begin{align*}
   \langle \vlambda, \vomega \rangle
     &= \sum_{i \in \setSc}
          (+L) \cdot \omega_i
        +
        \sum_{i \in \setS}
          (-L) \cdot \omega_i \\
     &\overset{\text{(a)}}{=}
        L
          \cdot
          \onenorm{\vomega_{\setSc}}
        -
        L
          \cdot
          \onenorm{\vomega_{\setS}}
      \overset{\text{(b)}}{>}
        0
      = \langle \vlambda, \vzero \rangle,
\end{align*}
where step~$\text{(a)}$ follows from the fact that $|\omega_i| = \omega_i$ for
all $i \in \setI(\matrHCC)$, and where step~$\text{(b)}$ follows
from~\eqref{eq:bsc:strict:balancedness:1}. Therefore, under \CCLPD\ the
all-zero codeword has the lowest cost function value when compared to all
non-zero pseudo-codewords in the fundamental cone, and therefore also compared
to all non-zero pseudo-codewords in the fundamental polytope.

\section{Proof of
               Lemma~\ref{lemma:meaning:of:bsc:pseudo:weight:1}}
\label{sec:proof:lemma:meaning:of:bsc:pseudo:weight:1}

\textbf{Case 1:} Let $\card{\setS} < \frac{1}{2} \cdot \wpsBSC(\vomega)$. The
proof is by contradiction: assume that $\onenorm{\vomega_{\setS}} \geq
\onenorm{\vomega_{\setSc}}$. This statement is clearly equivalent to the
statement that $2 \cdot \onenorm{\vomega_{\setS}} \geq
\onenorm{\vomega_{\setS}} + \onenorm{\vomega_{\setSc}} = \onenorm{\vomega}$,
which is equivalent to the statement that $\onenorm{\vomega_{\setS}} \geq
\frac{1}{2} \cdot \onenorm{\vomega}$. In terms of the notation in
Definition~\ref{def:pseudo:weights:1}, this means that
\begin{align*}
  \wpsBSC(\vomega)
    &= 2 \cdot F^{-1}\left( \frac{\onenorm{\vomega}}{2} \right)
     \overset{\text{(a)}}{\leq}
       2 \cdot F^{-1}(\onenorm{\vomega_{\setS}}) \\
    &\overset{\text{(b)}}{\leq}
       2
       \cdot
       \frac{\onenorm{\vomega_{\setS}}}{\infnorm{\vomega}}
     \leq
       2
       \cdot
       \frac{\card{\setS} \cdot \infnorm{\vomega}}{\infnorm{\vomega}}
     = 2
       \cdot
       \card{\setS},
\end{align*}
where at step~$\text{(a)}$ we have used the fact that $F^{-1}$ is a (strictly)
non-decreasing function and where at step~$\text{(b)}$ we have used the fact
that the slope of $F^{-1}$ (over the domain where $F^{-1}$ is defined) is at
least $1 / \infnorm{\vomega}$. The obtained inequality, however, is a
contradiction to the assumption that $\card{\setS} < \frac{1}{2} \cdot
\wpsBSC(\vomega)$.

\textbf{Case 2:} Let $\card{\setS} < \frac{1}{2} \cdot
\wpsBSCmod(\vomega)$. The proof is by contradiction: assume that
$\onenorm{\vomega_{\setS}} \geq \onenorm{\vomega_{\setSc}}$. Then, using the
definition of $\vomega'$ based on $\vomega$
(\confer~Section~\ref{sec:pseudo:weight:definitions:1}), we obtain
\begin{align*}
  \onenorm{\vomega'_{\{ 1, \ldots, \card{\setS} \}}}
    &\, \geq \,
       \onenorm{\vomega_{\setS}}
     \, \geq \,
       \onenorm{\vomega_{\setSc}}
     \, \geq \,
       \onenorm{\vomega'_{\{ \card{\setS}+1, \ldots, n \}}}.
\end{align*}
If $\wpsBSCmod(\vomega)$ is an even integer, then the above line of
inequalities shows that $\card{\setS} \geq \frac{1}{2} \cdot
\wpsBSCmod(\vomega)$, which is a contradiction to the assumption that
$\card{\setS} < \frac{1}{2} \cdot \wpsBSCmod(\vomega)$. If
$\wpsBSCmod(\vomega)$ is an odd integer, then the above line of inequalities
shows that $\card{\setS} \geq \frac{1}{2} \cdot \bigl(\wpsBSCmod(\vomega) + 1
\bigr) > \frac{1}{2} \wpsBSCmod(\vomega)$, which again is a contradiction to
the assumption that $\card{\setS} < \frac{1}{2} \cdot \wpsBSCmod(\vomega)$.

\section{Extensions of the Bridge Lemma}
\label{sec:extensions:bridge:lemma:1}

The aim of this appendix is to extend
Lemma~\ref{lemma:equation:nullspace:to:fc:1}
(\confer~Section~\ref{sec:bridge:1}) to measurement matrices beyond zero-one
matrices. In that vein we will present three generalizations in
Lemmas~\ref{lemma:equation:nullspace:to:fc:1:complex:case},
\ref{lemma:bridge:for:lifted:matrices:1},
and~\ref{lemma:equation:nullspace:to:fc:1:multiple:complex:vector:case}.  Note
that the setup in this appendix will be slightly more general than the
compressed sensing setup in Section~\ref{sec:cs:lpd:1} (and in most of the
rest of this paper). In particular, we allow matrices and vectors to be over
$\C$, and not just over~$\R$.

We will need some additional notation. Namely, similarly to the way that we
have extended the absolute value operator $\absnorm{\,\cdot\,}$ from scalars
to vectors at the beginning of Section~\ref{sec:bridge:1}, we will now extend
its use from scalars to matrices.

Moreover, we let $\Cnorm{\,\cdot\,}$ be an arbitrary norm for the complex
numbers. As such, $\Cnorm{\,\cdot\,}$ satisfies for any $a, b, c \in \C$ the
triangle inequality $\Cnorm{a+b} \leq \Cnorm{a} + \Cnorm{b}$ and the equality
$\Cnorm{c \cdot a} = \absnorm{c} \cdot \Cnorm{a}$. In the same way the
absolute value operator $\lvert\,\cdot\,\rvert$ was extended from scalars to
vectors and matrices, we extend the norm operator $\Cnorm{\,\cdot\,}$ from
scalars to vectors and matrices.

We let $\Cvectnorm{\,\cdot\,}$ be an arbitrary vector norm for complex vectors
that reduces to $\Cnorm{\,\cdot\,}$ for vectors with one component. As such,
$\Cvectnorm{\,\cdot\,}$ satisfies for any $c \in \C$ and any complex vectors
$\va$ and $\vb$ with the same number of components the triangle inequality
$\Cvectnorm{\va+\vb} \leq \Cvectnorm{\va} + \Cvectnorm{\vb}$ and the equality
$\Cvectnorm{c \cdot \va} = \absnorm{c} \cdot \Cvectnorm{\va}$.

We are now ready to discuss our first extension of
Lemma~\ref{lemma:equation:nullspace:to:fc:1}, which generalizes the setup of
that lemma from real measurement matrices where every entry is equal to either
zero or one to complex measurement matrices where the absolute value of every
entry is equal to either zero or one. Note that the upcoming lemma also
generalizes the mapping that is applied to the vectors in the nullspace of the
measurement matrix.

\begin{lemma}
  \label{lemma:equation:nullspace:to:fc:1:complex:case}
 
  Let $\matrHCS = (h_{j,i})_{j,i}$ be a measurement matrix over $\C$ such that
  $\absnorm{h_{j,i}} \in \{ 0, 1 \}$ for all $(j,i) \in \setJ(\matrHCS) \times
  \setI(\matrHCS)$, and let $\Cnorm{\,\cdot\,}$ be an arbitrary norm on
  $\C$. Then
  \begin{align*}
    \vnu \in \nullspaceC(\matrHCS)
    \ \ \ \Rightarrow \ \ \ 
    \Cnorm{\vnu} \in \fc{K}\big( \absnorm{\matrHCS} \big).
  \end{align*}
\end{lemma}

\noindent
\qquad \emph{Remark:} Note that $\supp(\vnu) = \supp(\Cnorm{\vnu})$.

\begin{IEEEproof}
  Let $\vomega \defeq \Cnorm{\vnu}$. In order to show that such a vector
  $\vomega$ is indeed in the fundamental cone of $\absnorm{\matrHCS}$, we need
  to verify~\eqref{eq:fund:cone:def:1} and~\eqref{eq:fund:cone:def:2}. The way
  $\vomega$ is defined, it is clear that it
  satisfies~\eqref{eq:fund:cone:def:1}. Therefore, let us focus on the proof
  that $\vomega$ satisfies~\eqref{eq:fund:cone:def:2}. Namely, from $\vnu \in
  \nullspaceC(\matrHCS)$ it follows that for all $j \in \setJ$, $\sum_{i \in
    \setI} h_{j,i} \nu_i = 0$. For all $j \in \setJ$ and all $i \in \setI_j$
  this implies that
  \begin{align*}
    \omega_i
      &= \Cnorm{\nu_i}
       = \absnorm{h_{j,i}} \cdot \Cnorm{\nu_i}
       = \Cnorm{h_{j,i} \nu_i}
       = \Cnorm{
           \,\, - \!\!
           \sum_{i' \in \setI \setminus i}
             h_{j,i'} \nu_{i'}
         } \\
      &\leq
         \sum_{i' \in \setI \setminus i}
            \Cnorm{h_{j,i'} \nu_{i'}}
       = \sum_{i' \in \setI \setminus i}
            \absnorm{h_{j,i'}} \cdot \Cnorm{\nu_{i'}}
       = \sum_{i' \in \setI_j \setminus i}
            \Cnorm{\nu_{i'}} \\
      &= \sum_{i' \in \setI_j \setminus i}
            \omega_{i'},
  \end{align*}
  showing that $\vomega$ indeed satisfies~\eqref{eq:fund:cone:def:2}.
\end{IEEEproof}

\mbox{}

\begin{example}
  The measurement matrix
  \begin{align*}
    \matrHCS
      &\defeq
         \begin{pmatrix}
            1 & 0 & \frac{1}{\sqrt{2}}(1+i) \\
           -1 & i & 1
         \end{pmatrix}
  \end{align*}
  satisfies
  \begin{align*}
    \absnorm{\matrHCS}
      &= \begin{pmatrix}
            1 & 0 & 1 \\
            1 & 1 & 1
         \end{pmatrix},
  \end{align*}
  and so Lemma~\ref{lemma:equation:nullspace:to:fc:1:complex:case} is
  applicable. An example of a vector in $\nullspaceC(\matrHCS)$ is
  \begin{align*}
    \vnu
      &\defeq
         \left(
           \frac{1}{\sqrt{2}}(1+i), \ 
           \frac{1}{\sqrt{2}} - i\left( 1+\frac{1}{\sqrt{2}} \right), \ 
           -1
         \right).
  \end{align*}
  Choosing $\Cnorm{\,\cdot\,} \defeq \absnorm{\,\cdot\,}$, we obtain
  \begin{align*}
    \Cnorm{\vnu}
      &= \left(
           1, \ 
           \sqrt{2 + \sqrt{2}}, \ 
           1
         \right)
       = \left(
           1, \ 
           1.848..., \ 
           1
         \right)
       \in \fc{K}\big( \absnorm{\matrHCS} \big).
  \end{align*}
\end{example}

The second extension of Lemma~\ref{lemma:equation:nullspace:to:fc:1}
generalizes that lemma to hold also for complex measurement matrices where the
absolute value of every entry is an integer. In order to present this lemma,
we need the following definition, which is subsequently illustrated by
Example~\ref{example:HCS:cover:1}.

\begin{definition}
  \label{def:measurement:matrix:cover:1}

  Let $\matrHCS = (h_{j,i})_{j, i}$ be a measurement matrix over $\C$ such
  that $\absnorm{h_{j,i}} \in \Zp$ for all $(j,i) \in \setJ(\matrHCS) \times
  \setI(\matrHCS)$, and let $M \in \Zpp$ be such that $M \geq \max_{(j,i)} \,
  \absnorm{h_{j,i}}$. We define an $M$-fold cover $\tmatrHCS$ of $\matrHCS$ as
  follows: for $(j,i) \in \setJ(\matrHCS) \times \setI(\matrHCS)$, if the
  scalar $h_{j,i}$ is non-zero then it is replaced by a matrix, namely
  $h_{j,i} / \absnorm{h_{j,i}}$ times the sum of $\absnorm{h_{j,i}}$ arbitrary
  $M \times M$ permutation matrices with non-overlapping support. However, if
  $h_{j,i} = 0$ then the scalar $h_{j,i}$ is replaced by an all-zero matrix of
  size $M \times M$.
\end{definition}

\mbox{}

Note that all entries of the matrix $\tmatrHCS$ in
Definition~\ref{def:measurement:matrix:cover:1} have absolute value equal to
either zero or one.

\mbox{}

\begin{example}
  \label{example:HCS:cover:1}

  Let
  \begin{align*}
    \matrHCS
      &\defeq
         \begin{pmatrix}
            1 & 0 & \sqrt{2}(1+i) \\
           -2 & i & 3
         \end{pmatrix}.
  \end{align*}
  Clearly
  \begin{align*}
    \absnorm{\matrHCS}
      &= \begin{pmatrix}
            1 & 0 & 2 \\
            2 & 1 & 3
         \end{pmatrix},
  \end{align*}
  and so, choosing $M \defeq 3$ and
  \begin{align*}
    \tmatrHCS
      &\defeq
         \left(
           \begin{array}{ccc|ccc|ccc}
              0 & 1 & 0 & 0 & 0 & 0 & \frac{1+i}{\sqrt{2}} & \frac{1+i}{\sqrt{2}} & 0 \\
              1 & 0 & 0 & 0 & 0 & 0 & \frac{1+i}{\sqrt{2}} & 0 & \frac{1+i}{\sqrt{2}} \\
              0 & 0 & 1 & 0 & 0 & 0 & 0 & \frac{1+i}{\sqrt{2}} & \frac{1+i}{\sqrt{2}} \\
              \hline
               0 & -1 & -1 & i & 0 & 0 & 1 & 1 & 1 \\
              -1 & -1 &  0 & 0 & i & 0 & 1 & 1 & 1 \\
              -1 &  0 & -1 & 0 & 0 & i & 1 & 1 & 1 \\
           \end{array}
         \right),
  \end{align*}
  we obtain a matrix described by the procedure of
  Definition~\ref{def:measurement:matrix:cover:1}.
\end{example}

\begin{lemma}
  \label{lemma:bridge:for:lifted:matrices:1}

  Let $\matrHCS = (h_{j,i})_{j,i}$ be a measurement matrix over $\C$ such that
  $\absnorm{h_{j,i}} \in \Zp$ for all $(j,i) \in \setJ(\matrHCS) \times
  \setI(\matrHCS)$. Let $M \in \Zpp$ be such that $M \geq \max_{(j,i)} \,
  \absnorm{h_{j,i}}$, and let $\tmatrHCS$ be a matrix obtained by the
  procedure in Definition~\ref{def:measurement:matrix:cover:1}. Moreover, let
  $\Cnorm{\,\cdot\,}$ be an arbitrary norm on $\C$. Then
  \begin{align*}
    \vnu \in \nullspaceC(\matrHCS)
    & \ \ \ \Rightarrow \ \ \ 
    \vnu^{\uparrow M} \in \nullspaceC(\tmatrHCS) \\
    & \ \ \ \Rightarrow \ \ \ 
    \Cnorm{\vnu^{\uparrow M}} \in \fc{K}\big( \absnorm{\tmatrHCS} \big).
  \end{align*}
  Additionally, with respect to the first implication sign we have the
  following converse: for any $\tvnu \in \C^{Mn}$ we have
  \begin{align*}
    \varphiM(\tvnu) \in \nullspaceC(\matrHCS)
    & \ \ \ \Leftarrow \ \ \ 
    \tvnu \in \nullspaceC(\tmatrHCS).
  \end{align*}
\end{lemma}

\begin{IEEEproof}
  Let $\tmatrHCS = (\tilde h_{(j,m'),(i,m)})_{(j,m'),(i,m)}$. Note that by the
  construction in Definition~\ref{def:measurement:matrix:cover:1}, it holds
  that
  \begin{align*}
    \sum_{m' \in \set [M]}
      \tilde h_{(j,m'),(i,m)}
      &= h_{j,i}
      \ \ \text{for any $(j,i,m) \in \setJ \!\times\! \setI \!\times\! [M]$}, \\
    \sum_{m \in \set [M]}
      \tilde h_{(j,m'),(i,m)}
      &= h_{j,i}
      \ \ \text{for any $(j,m',i) \in \setJ \!\times\! [M] \!\times\! \setI$}.
  \end{align*}
  Let $\vnu \in \nullspaceC(\matrHCS)$. Then, for every $(j,m') \in \setJ
  \times [M]$ we have
  \begin{align*}
     &
     \sum_{(i,m) \in \setI \times \set [M]} \!\!\!
      \tilde h_{(j,m'),(i,m)} \nu^{\uparrow M}_{(i,m)}
       =  \sum_{(i,m) \in \setI \times \set [M]} \!\!\!
           \tilde h_{(j,m'),(i,m)}
           \nu_i \\
      &\quad\quad
       =  \sum_{i \in \setI}
           \nu_i
           \sum_{m \in \set [M]}
           \tilde h_{(j,m'),(i,m)}
       =  \sum_{i \in \setI}
           \nu_i
           h_{j,i}
       = 0,
  \end{align*}
  where the last equality follows from the assumption that $\vnu \in
  \nullspaceC(\matrHCS)$. Therefore $\vnu^{\uparrow M} \in
  \nullspaceC(\tmatrHCS)$. Because $\absnorm{\tilde h_{(j,m'),(i,m)}} \in \{
  0, 1 \}$ for all $(j,m',i,m) \in \setJ \times [M] \times \setI \times [M]$,
  we can then apply Lemma~\ref{lemma:equation:nullspace:to:fc:1:complex:case}
  to conclude that $\Cnorm{\vnu^{\uparrow M}} \in \fc{K}\big(
  \absnorm{\tmatrHCS} \big)$.
  
  Now, in order to prove the last part of the lemma, assume that $\tvnu \in
  \nullspaceC(\tmatrHCS)$ and define $\vnu \defeq \varphiM(\tvnu)$. Then for
  every $j \in \setJ$ we have
  \begin{align*}
    \sum_{i \in \setI}
           h_{j,i}
           \nu_i
      &= \sum_{i \in \setI}
           h_{j,i}
           \cdot
           \frac{1}{M}
           \sum_{m \in [M]}
             \tilde \nu_{(i,m)} \\
      &= \frac{1}{M}
         \sum_{i \in \setI}
         \sum_{m \in [M]}
           h_{j,i}
           \cdot
             \tilde \nu_{(i,m)} \\
      &= \frac{1}{M}
         \sum_{i \in \setI}
         \sum_{m \in [M]}
         \sum_{m' \in [M]}
           \tilde h_{(j,m'),(i,m)}
           \cdot
             \tilde \nu_{(i,m)} \\
      &= \frac{1}{M}
         \sum_{m' \in [M]}
         \left(
           \sum_{i \in \setI}
           \sum_{m \in [M]}
             \tilde h_{(j,m'),(i,m)}
             \cdot
               \tilde \nu_{(i,m)}
         \right) \\
      &= 0,
  \end{align*}
  where the last equality follows from the assumption that $\tvnu \in
  \nullspaceC(\tmatrHCS)$, \ie, for every $(j,m') \in \setJ \times [M]$ the
  expression in parentheses equals zero. Therefore, $\vnu = \varphiM(\tvnu)
  \in \nullspaceC(\matrHCS)$.
\end{IEEEproof}

\mbox{}

\begin{example}
  Consider the measurement matrix $\matrHCS$ of
  Example~\ref{example:HCS:cover:1}. A possible vector in
  $\nullspaceC(\matrHCS)$ is given by
  \begin{align*}
    \vnu
      &\defeq
         \left(
           \sqrt{2}(1+i), \ 
           2\sqrt{2} - i \big( 3 + 2\sqrt{2} \big), \ 
           -1
         \right).
  \end{align*}
  Applying Lemma~\ref{lemma:bridge:for:lifted:matrices:1} with $M \defeq 3$ and
  $\Cnorm{\,\cdot\,} \defeq\absnorm{\,\cdot\,}$, we obtain
  \begin{align*}
    \Cnorm{\vnu^{\uparrow 3}}
      &= \left(
           2,
           2,
           2, 
           \alpha,
           \alpha,
           \alpha, \ 
           1,
           1,
           1
         \right)
           \in \fc{K}\big( \absnorm{\tmatrHCS} \big),
  \end{align*}
  where $\alpha = \sqrt{25 + 12\sqrt{2}} = 6.478...\,$, and where $\tmatrHCS$
  can be chosen as in Example~\ref{example:HCS:cover:1}.
\end{example}

Our third extension of Lemma~\ref{lemma:equation:nullspace:to:fc:1}
generalizes the mapping that is applied to the vectors in the nullspace of the
measurement matrix.

\begin{lemma}
  \label{lemma:equation:nullspace:to:fc:1:multiple:complex:vector:case}

  Let $\matrHCS = (h_{j,i})_{j,i}$ be a measurement matrix over $\C$ such that
  $\absnorm{h_{j,i}} \in \{ 0, 1 \}$ for all $(j,i) \in \setJ(\matrHCS) \times
  \setI(\matrHCS)$. Let $L \in \Zpp$, let $\Cvectnorm{\,\cdot\,}$ be an
  arbitrary norm for complex vectors, and let $\{ \vnu^{(\ell)} \}_{\ell \in
    [L]}$ be a collection of vectors with $n$ components. Then
  \begin{align*}
    \vnu^{(1)}, \ldots, \vnu^{(L)} \in \nullspaceC(\matrHCS)
    \ \ \Rightarrow \ \ 
    \vomega \in \fc{K}\big( \absnorm{\matrHCS} \big),
  \end{align*}
  where $\vomega \in \R^n$ is defined such that for all $i \in
  \setI(\matrHCS)$,
  \begin{align*}
    \omega_i
      &= \Cvectnorm{\left( \nu^{(1)}_i, \ldots, \nu^{(L)}_i \right)}.
  \end{align*}
\end{lemma}

\begin{IEEEproof}
  The proof is very similar to the proof of
  Lemma~\ref{lemma:equation:nullspace:to:fc:1:complex:case}. Namely, in order
  to show that $\vomega$ is indeed in the fundamental cone of
  $\absnorm{\matrHCS}$, we need to verify~\eqref{eq:fund:cone:def:1}
  and~\eqref{eq:fund:cone:def:2}. The way $\vomega$ is defined, it is clear
  that it satisfies~\eqref{eq:fund:cone:def:1}. Therefore, let us focus on the
  proof that $\vomega$ satisfies~\eqref{eq:fund:cone:def:2}. Namely, from
  $\vnu^{(\ell)} \in \nullspaceC(\matrHCS)$, $\ell \in [L]$, it follows that
  $\sum_{i \in \setI} h_{j,i} \nu^{(\ell)}_i = 0$, $j \in \setJ$, $\ell \in
  [L]$. For all $j \in \setJ$ and all $i \in \setI_j$ this implies that
  \begin{align*}
    \omega_i
      &= \Cvectnorm{\left( \nu^{(1)}_i, \ldots, \nu^{(L)}_i \right)} \\
      &= \absnorm{h_{j,i}}
         \cdot
         \Cvectnorm{\left( \nu^{(1)}_i, \ldots, \nu^{(L)}_i \right)} \\
      &= \Cvectnorm{
           \left(
             h_{j,i} \nu^{(1)}_i, \ldots, h_{j,i} \nu^{(L)}_i
           \right)} \\
      &= \Cvectnorm{
           \left(
             \,\, - \!\!
             \sum_{i' \in \setI \setminus i}
               h_{j,i'} \nu^{(1)}_{i'},
             \ldots,
             \,\, - \!\!
             \sum_{i' \in \setI \setminus i}
               h_{j,i'} \nu^{(L)}_{i'}
           \right)
         } \\
      &= \Cvectnorm{
           \,\, - \!\!
           \sum_{i' \in \setI \setminus i}
             h_{j,i'} 
             \cdot
             \left(
               \nu^{(1)}_{i'},
               \ldots,
               \nu^{(L)}_{i'}
             \right)
         } \\
      &\leq
         \sum_{i' \in \setI \setminus i}
         \Cvectnorm{
           h_{j,i'} 
           \cdot
           \left(
             \nu^{(1)}_{i'},
             \ldots,
             \nu^{(L)}_{i'}
           \right)
         } \\
      &= \sum_{i' \in \setI \setminus i}
         \absnorm{h_{j,i'}} 
         \cdot
         \Cvectnorm{
           \left(
             \nu^{(1)}_{i'},
             \ldots,
             \nu^{(L)}_{i'}
           \right)
         } \\
      &= \sum_{i' \in \setI_j \setminus i}
           \Cvectnorm{\left( \nu^{(1)}_{i'}, \ldots, \nu^{(L)}_{i'} \right)} \\
      &= \sum_{i' \in \setI_j \setminus i}
            \omega_{i'},
  \end{align*}
  showing that $\vomega$ indeed satisfies~\eqref{eq:fund:cone:def:2}.
\end{IEEEproof}

\mbox{}

\begin{corollary}
  \label{cor:lemma:equation:nullspace:to:fc:1:multiple:complex:vector:case}

  Consider the setup of
  Lemma~\ref{lemma:equation:nullspace:to:fc:1:multiple:complex:vector:case}.
  Let $L \in \Zpp$, and select $L$ arbitrary scalars $\alpha^{(\ell)} \in
  \Rp$, $\ell \in [L]$, and $L$ arbitrary vectors $\vnu^{(\ell)} \in
  \nullspaceC(\matrHCS)$, $\ell \in [L]$.
  \begin{itemize}

  \item For $\Cvectnorm{\,\cdot\,} \defeq \onenorm{\,\cdot\,}$ we have
    \begin{align*}
      \sum_{\ell \in [L]}
        \alpha^{(\ell)} \,
        \absnorm{\vnu^{(\ell)}}
        &\in \fc{K}\big( \absnorm{\matrHCS} \big).
    \end{align*}

  \item For $\Cvectnorm{\,\cdot\,} \defeq \twonorm{\,\cdot\,}$ we have
    \begin{align*}
      \sqrt{
        \sum_{\ell \in [L]}
          (\alpha^{(\ell)})^2 \,
          \absnorm{\vnu^{(\ell)}}^2
      }
        &\in \fc{K}\big( \absnorm{\matrHCS} \big),
    \end{align*}
    where the square root and the square of a vector are understood
    component-wise.

  \end{itemize}
\end{corollary}

\begin{IEEEproof}
  These are straightforward consequences of applying
  Lemma~\ref{lemma:equation:nullspace:to:fc:1:multiple:complex:vector:case} to
  $\big\{ \alpha^{(\ell)} \cdot \vnu^{(\ell)} \big\}_{\ell \in [L]}$.
\end{IEEEproof}

\mbox{}

Because $\fc{K}\big( \absnorm{\matrHCS} \big)$ is a convex cone, the first
statement in
Corollary~\ref{cor:lemma:equation:nullspace:to:fc:1:multiple:complex:vector:case}
can also be proven by combining $\absnorm{\vnu^{(\ell)}} \in \fc{K}\big(
\absnorm{\matrHCS} \big)$, $\ell \in [L]$, with the fact that any conic
combination of vectors in $\fc{K}\big( \absnorm{\matrHCS} \big)$ is a vector
in $\fc{K}\big( \absnorm{\matrHCS} \big)$. In that respect, the second
statement of
Corollary~\ref{cor:lemma:equation:nullspace:to:fc:1:multiple:complex:vector:case}
is noteworthy in the sense that although $L$ vectors in $\fc{K}\big(
\absnorm{\matrHCS} \big)$ are combined in a ``non-conic'' way, we nevertheless
obtain a vector in $\fc{K}\big( \absnorm{\matrHCS} \big)$. (Of course, for the
latter to work it is important that these $L$ vectors are not arbitrary
vectors in $\fc{K}\big( \absnorm{\matrHCS} \big)$ but that they are derived
from vectors in the $\C$-nullspace of $\matrHCS$.)

We conclude this appendix with two remarks. First, it is clear that
Lemma~\ref{lemma:equation:nullspace:to:fc:1:multiple:complex:vector:case} can
be extended in the same way as Lemma~\ref{lemma:bridge:for:lifted:matrices:1}
extends Lemma~\ref{lemma:equation:nullspace:to:fc:1:complex:case}. Second,
although most of Section~\ref{sec:translation:1} is devoted to using
Lemma~\ref{lemma:equation:nullspace:to:fc:1} for translating ``positive
results'' about \CCLPD\ to ``positive results'' about \CSLPD\ , it is clear
that Lemmas~\ref{lemma:equation:nullspace:to:fc:1:complex:case},
\ref{lemma:bridge:for:lifted:matrices:1},
and~\ref{lemma:equation:nullspace:to:fc:1:multiple:complex:vector:case} can
equally well be the basis for translating results from \CCLPD\ to \CSLPD.

\section{Proof of
               Theorem~\ref{theorem:weight:l2:l1:and:approximation:guarantees:1}}
\label{sec:proof:theorem:weight:l2:l1:and:approximation:guarantees:1}

By definition, $\ve$ is the original signal. Since $\matrHCS \cdot \ve =
\vs$ and $\matrHCS \cdot \hve = \vs$, it easily follows that $\vnnu \defeq \ve
- \hve$ is in the nullspace of $\matrHCS$. So,
\begin{align}
  \!\!\!\!\!
  \onenorm{\veS}
  +
  \onenorm{\veSc}
    &= \onenorm{\ve} \label{eq:ell2:ell1:guarantee:1:start} \\
    &\overset{\text{(a)}}{\geq}
       \onenorm{\hve} \nonumber \\
    &= \onenorm{\ve - \vnnu} \nonumber \\
    &= \onenorm{\veS - \vnnuS}
       + 
       \onenorm{\veSc - \vnnuSc} \nonumber \\
    &\overset{\text{(b)}}{\geq}
       \onenorm{\veS}
       -
       \onenorm{\vnnuS}
       +
       \onenorm{\vnnuSc}
       -
       \onenorm{\veSc} \label{eq:ell2:ell1:guarantee:1:before:end} \\
    &\overset{\text{(c)}}{\geq}
       \onenorm{\veS}
       \!+\!
       \left(
         \!
         \sqrt{C'}
         \!-\!
         2\sqrt{k}
       \right) \!
       \twonorm{\vnnu}
       \!-\!
       \onenorm{\veSc},
         \label{eq:ell2:ell1:guarantee:1}
\end{align}
where step~$\text{(a)}$ follows from the fact that the solution of \CSLPD\
satisfies $\onenorm{\hve} \leq \onenorm{\ve}$ and where step~$\text{(b)}$
follows from applying the triangle inequality property of the $\ell_1$-norm
twice. Moreover, step~$\text{(c)}$ follows from
\begin{align*}
  -
  \onenorm{\vnnuS}
  +
  \onenorm{\vnnuSc}
    &= \onenorm{\vnnu}
       -
       2
         \onenorm{\vnnuS} \\
    &\overset{\text{(d)}}{\geq}
       \sqrt{C'}
         \twonorm{\vnnu}
       -
       2
         \onenorm{\vnnuS} \\
    &\overset{\text{(e)}}{\geq}
       \sqrt{C'}
         \twonorm{\vnnu}
       -
       2
         \sqrt{k}
         \twonorm{\vnnuS} \\
    &\overset{\text{(f)}}{\geq}
       \sqrt{C'}
         \twonorm{\vnnu}
       -
       2
         \sqrt{k}
         \twonorm{\vnnu} \\
    &= \left(
         \sqrt{C'}
         -
         2\sqrt{k}
       \right)
       \twonorm{\vnnu},
\end{align*}
where step~$\text{(d)}$ follows from the assumption that $\wpsAWGNC(|\vnnu|)
\geq C'$ holds for all $\vnnu \in \nullspaceR(\matrHCS) \setminus \{ \vzero
\}$, \ie, that $\onenorm{\vnnu} \geq \sqrt{C'} \cdot \twonorm{\vnnu}$ holds
for all $\vnnu \in \nullspaceR(\matrHCS)$, where step~$\text{(e)}$ follows
from the inequality $\onenorm{\va} \leq \sqrt{k} \cdot \twonorm{\va}$ that
holds for any real vector $\va$ with $k$ components, and where
step~$\text{(f)}$ follows from the inequality $\twonorm{\va_{\set{S}}} \leq
\twonorm{\va}$ that holds for any real vector $\va$ whose set of coordinate
indices includes $\set{S}$. Subtracting the term $\onenorm{\veS}$ on both
sides
of~\eqref{eq:ell2:ell1:guarantee:1:start}--\eqref{eq:ell2:ell1:guarantee:1},
and solving for $\twonorm{\vnnu} = \twonorm{\ve - \hve}$, we obtain the claim.

\section{Proof of
               Theorem~\ref{theorem:weight:linfty:l1:and:approximation:guarantees:1}}
\label{sec:proof:theorem:weight:linfty:l1:and:approximation:guarantees:1}

By definition, $\ve$ is the original signal. Since $\matrHCS
\cdot \ve = \vs$ and $\matrHCS \cdot \hve = \vs$, it easily follows that
$\vnnu \defeq \ve - \hve$ is in the nullspace of $\matrHCS$. So,
\begin{align}
  \!\!\!\!
  \onenorm{\veS}
  +
  \onenorm{\veSc}
    &= \onenorm{\ve} \label{eq:ellinfty:ell1:guarantee:1:start} \\
    &\overset{\text{(a)}}{\geq}
       \onenorm{\veS}
       -
       \onenorm{\vnnuS}
       +
       \onenorm{\vnnuSc}
       -
       \onenorm{\veSc} \nonumber \\
    &\overset{\text{(b)}}{\geq}
       \onenorm{\veS}
       +
       \left(
         C'
         -
         2k
       \right)
       \cdot
       \infnorm{\vnnu}
       -
       \onenorm{\veSc},
         \label{eq:ellinfty:ell1:guarantee:1}
\end{align}
where step~$\text{(a)}$ follows from the same line of reasoning as in going
from~\eqref{eq:ell2:ell1:guarantee:1:start}
to~\eqref{eq:ell2:ell1:guarantee:1:before:end}, and where step~$\text{(b)}$
follows from
\begin{align*}
  -
  \onenorm{\vnnuS}
  +
  \onenorm{\vnnuSc}
    &= \onenorm{\vnnu}
       -
       2
         \cdot
         \onenorm{\vnnuS} \\
    &\overset{\text{(c)}}{\geq}
       C'
         \cdot
         \infnorm{\vnnu}
       \!-\!
       2
         \cdot
         \onenorm{\vnnuS} \\
    &\overset{\text{(d)}}{\geq}
       C'
         \cdot
         \infnorm{\vnnu}
       \!-\!
       2k
         \cdot
         \infnorm{\vnnuS} \\
    &\overset{\text{(e)}}{\geq}
       C'
         \cdot
         \infnorm{\vnnu}
       -
       2k
         \cdot
         \infnorm{\vnnu} \\
    &= \left(
         C'
         -
         2k
       \right)
       \cdot
       \infnorm{\vnnu},
\end{align*}
where step~$\text{(c)}$ follows from the assumption that $\wmaxfr(|\vnnu|)
\geq C'$ holds for all $\vnnu \in \nullspaceR(\matrHCS) \setminus \{ \vzero
\}$, \ie, $\onenorm{\vnnu} \geq C' \cdot \infnorm{\vnnu}$ holds for all $\vnnu
\in \nullspaceR(\matrHCS)$, where step $\text{(d)}$ follows from the
inequality $\onenorm{\va} \leq k \cdot \infnorm{\va}$ that holds for any real
vector $\va$ with $k$ components, and where step~$\text{(e)}$ follows the
inequality $\infnorm{\va_{\set{S}}} \leq \infnorm{\va}$ that holds for any
real vector $\va$ whose set of coordinate indices includes
$\set{S}$. Subtracting the term $\onenorm{\veS}$ on both sides
of~\eqref{eq:ellinfty:ell1:guarantee:1:start}--\eqref{eq:ellinfty:ell1:guarantee:1},
and solving for $\infnorm{\vnnu} = \infnorm{\ve - \hve}$ we obtain the claim.

\section{Proof of
               Theorem~\ref{theorem:mld:reformulation:1}}
\label{sec:proof:theorem:mld:reformulation:1}

In a first step, we discuss the reformulation of the cost function. Namely,
for arbitrary $\vx' \in \codeCCC$, let $\ve' \defeq \ovy - \vx' \ \inGFtwo$,
\ie, $x'_i = \oy_i - e'_i = \oy_i + e'_i \ \inGFtwo$ for all $i \in
\setI$. Then
\begin{align}
  \sum_{i \in \setI}
    \lambda_i x'_i
    &\overset{\text{(a)}}{=}
       \sum_{i \in \setI}
         \lambda_i (\oy_i + e'_i - 2\oy_i e'_i) \nonumber \\
    &= \sum_{i \in \setI}
         \lambda_i \oy_i
       +
       \sum_{i \in \setI}
         \lambda_i \cdot (1 - 2\oy_i) \cdot e'_i \nonumber \\
    &\overset{\text{(b)}}{=}
       \sum_{i \in \setI}
         \lambda_i \oy_i
       +
       \sum_{i \in \setI}
         \absnorm{\lambda_i}
         \cdot
         e'_i,
           \label{eq:inner:product:reformulation:1}
\end{align}
where at step $\text{(a)}$ we used the fact that for $a, b \in \{0, 1 \}$, the
result of $a + b \ \inGFtwo$ can be written over the reals as $a + b - 2 a b$,
and at step $\text{(b)}$ we used the fact that for all $i \in \setI$,
$\lambda_i \cdot (1-2\oy_i) = \absnorm{\lambda_i}$. Notice that the first sum
in the last line of~\eqref{eq:inner:product:reformulation:1} is only a
function of $\vy$, hence minimizing $\langle \vlambda, \vx' \rangle = \sum_{i
  \in \setI} \lambda_i x'_i$ over $\vx'$ is equivalent to minimizing $\sum_{i
  \in \setI} \absnorm{\lambda_i} \cdot e'_i = \langle \absnorm{\vlambda}, \ve'
\rangle = \onenorm{\vlambda_{\supp(\ve')}}$ over $\ve'$.

In a second step, we discuss the reformulation of the constraint. Namely, for
arbitrary $\vx' \in \codeCCC$, and corresponding $\ve' \defeq \ovy - \vx' \
\inGFtwo$, we have $\matrHCC \cdot \ve' = \matrHCC \cdot (\ovy-\vx') =
\matrHCC \cdot \ovy - \matrHCC \cdot \vx' = \matrHCC \cdot \ovy - \vzero = \vs
\ \inGFtwo$.

\section{Proof of
               Theorem~\ref{theorem:CSLPD:reformulation:2:vs:1}}
\label{sec:proof:theorem:CSLPD:reformulation:2:vs:1}

Because for $M = 1$ the measurement matrix $\tmatrHCS$ equals the
measurement matrix $\matrHCS$, it is clear that any feasible vector of
\CSLPD\ yields a feasible vector of \CSLPDver{1}.

Therefore, let us show that for $M > 1$ no feasible vector of \CSLPDver{1}
yields a smaller cost function value than the cost function value of the
best feasible vector in the base Tanner graph. To that end, we demonstrate
that for any $M \in \Zpp$, any $M$-cover based $\tmatrHCS$, and any $\tve'$
with $\tmatrHCS \cdot \tve' = \vs^{\uparrow M}$, the cost function value of
$\tve'$ is never smaller than the cost function value of the feasible vector
in the base Tanner graph given by the projection $\varphiM(\tve')$. Indeed,
the cost function value of $\varphiM(\tve')$ is
\begin{align*}
  \onenorm{\varphiM(\tve')}
    &= \sum_{i \in \setI}
         \Absnorm{
           \frac{1}{M}
           \sum_{m \in [M]}\tilde e'_{i,m}
         }
     \leq
        \sum_{i \in \setI}
          \frac{1}{M}
          \sum_{m \in [M]}
            \absnorm{\tilde e'_{i,m}} \\
    &= \frac{1}{M}
       \sum_{i \in \setI}
         \sum_{m \in [M]}
           \absnorm{\tilde e'_{i,m}}
     = \frac{1}{M}
       \cdot
       \onenorm{\tve'},
\end{align*}
\ie, it is never larger than the cost function value of $\tve'$. Moreover,
since $\tmatrHCS \cdot \tve' = \vs^{\uparrow M}$ implies that $\matrHCS \cdot
\varphiM(\tve') = \vs$, we have proven the claim that $\varphiM(\tve') = \vs$
is a feasible vector in the base Tanner graph.

\section{Proof of
               Theorem~\ref{theorem:CSLPD:and:CSOPTzeroinfty:connection:1}}
\label{sec:proof:theorem:CSLPD:and:CSOPTzeroinfty:connection:1}

The proof has two parts. First we show that the minimal cost function value of
$\CSRELzeroinf$ is never smaller than the minimal cost function value of
\CSLPD. Second, we show that for any vector that minimizes the cost function
of $\CSLPD$ there is a graph cover and a configuration therein whose
zero-infinity operator equals the minimal cost function value of \CSLPD.

We prove the first part. Let $\ve'$ minimize $\onenorm{\ve'}$ over all $\ve'$
such that $\matrHCS \cdot \ve' = \vs$. For any $M \in \Zpp$, any $\tmatrHCS$
whose Tanner graph is an $M$-cover of the Tanner graph of $\matrHCS$, and any
$(\tve',\tvs)$ with $\tmatrHCS \cdot \tve' = \tvs$ and $\varphiM(\tvs) = \vs$,
it holds that
\begin{align*}
  \frac{1}{M}
  \zeroinfoperator{\tve'}
    &\overset{\text{(a)}}{\geq}
       \frac{1}{M}
       \onenorm{\tve'}
     \overset{\text{(b)}}{\geq}
       \onenorm{\varphiM(\tve')}
     \overset{\text{(c)}}{\geq}
       \onenorm{\ve'},
\end{align*}
where step $\text{(a)}$ follows from
Lemma~\ref{lemma:onenorm:zeroinfoperator:relationship:1}, where step
$\text{(b)}$ uses the same line of reasoning as the proof of
Theorem~\ref{theorem:CSLPD:reformulation:2:vs:1}, and where step $\text{(c)}$
follows from the easily verified fact that $\matrHCS \cdot \varphiM(\tve') =
\vs$, along with the definition of $\ve'$. Because $(\tve',\tvs)$ was
arbitrary (subject to $\tmatrHCS \cdot \tve' = \tvs$ and $\varphiM(\tvs) =
\vs$), this observation concludes the first part of the proof.

We now prove the second part. Again, let $\ve'$ minimize $\onenorm{\ve'}$ over
all $\ve'$ such that $\matrHCS \cdot \ve' = \vs$. Once \CSLPD\ is rewritten as
a linear program (with the help of suitable auxiliary variables), we see that
the coefficients that appear in this linear program are all rationals. Using
Cram\'er's rule for determinants, it follows that the set of feasible points
of this linear program is a polyhedral set whose vertices are all vectors with
rational entries. Therefore, if $\ve'$ is unique then $\ve'$ is a vector with
rational entries. If $\ve'$ is not unique then there is at least one vector
$\ve'$ with rational entries that minimizes the cost function of \CSLPD. Let
$\ve'$ be such a vector.

Before continuing, let us simplify the notation slightly. Namely, we
rearrange the constraint $\matrHCS \cdot \ve' = \vs$ in $\CSLPD$ so that it
reads
\begin{align}
  \begin{pmatrix}
    \matrHCS & -\matr{I}
  \end{pmatrix}
  \cdot
  \begin{pmatrix}
    \ve' \\
    \vs
  \end{pmatrix}
    &= \vzero,
         \label{eq:CSLPD:constraint:1}
\end{align}
and then we replace~\eqref{eq:CSLPD:constraint:1} by
\begin{align*}
  \matrHCS \cdot \ve'
    &= \vzero.
\end{align*}
This is done by redefining $\matrHCS$ to stand for $\bigl( \matrHCS, \
-\matr{I} \bigr)$, and redefining $\ve'$ to stand for $\bigl( (\ve')^\tr, \
(\vs)^\tr \bigr)^\tr$. Note that the redefined $\matrHCS$ contains zeros,
ones, or minus ones. Similarly, we rearrange the constraint $\tmatrHCS \cdot
\tve' = \tvs$ in $\CSRELzeroinf$ so that it reads
\begin{align}
  \begin{pmatrix}
    \tmatrHCS & -\matr{I}
  \end{pmatrix}
  \cdot
  \begin{pmatrix}
    \tve' \\
    \tvs
  \end{pmatrix}
    &= \vzero,
         \label{eq:CSRELzeroinf:constraint:1}
\end{align}
and then we replace~\eqref{eq:CSRELzeroinf:constraint:1} by
\begin{align*}
  \tmatrHCS \cdot \tve'
    &= \vzero.
\end{align*}
This is done by redefining $\tmatrHCS$ to stand for $\bigl( \tmatrHCS, \
-\matr{I} \bigr)$, and redefining $\tve'$ to stand for $\bigl( (\tve')^\tr, \
(\tvs)^\tr \bigr)^\tr$. Note that the redefined $\tmatrHCS$ contains only
zeros, ones, or minus ones, and that the Tanner graph representing the
redefined $\tmatrHCS$ is a valid $M$-fold cover of the Tanner graph
representing the redefined $\matrHCS$.

We will now exhibit a suitable $M$-fold cover and a configuration $\tve'$
therein such that $\varphiM(\tve') = \ve'$ and such that for some $\gamma
\in \Rpp$ the vector $\tve'$ will satisfy
\begin{align}
  \tilde e'_{(i,m)}
    &\in
       \begin{cases}
         \{ 0, +\gamma \} & \text{if $e'_i > 0$} \\
         \{ 0 \}          & \text{if $e'_i = 0$} \\
         \{ 0, -\gamma \} & \text{if $e'_i < 0$}
       \end{cases},
       \quad
       (i,m) \in \setI \times [M].
         \label{eq:CSRELzeroinf:graph:cover:assignment:1}
\end{align}
Then for such a vector the following holds
\begin{align*}
  \frac{1}{M}
  \zeroinfoperator{\tve'}
    &\overset{\text{(a)}}{=}
       \frac{1}{M}
       \onenorm{\tve'}
     \overset{\text{(b)}}{=}
       \onenorm{\varphiM(\tve')}
     \overset{\text{(c)}}{=}
       \onenorm{\ve'},
\end{align*}
where step $\text{(a)}$ follows from the fact that the equality condition in
Lemma~\ref{lemma:onenorm:zeroinfoperator:relationship:1} is satisfied, step
$\text{(b)}$ follows from the fact that for every $i \in \setI$, all $\{
\tilde e'_{(i,m)} \}_{m \in [M], \, \tilde e'_{(i,m)} \neq 0}$ have the same
sign, and step $\text{(c)}$ follows from $\varphiM(\tve') = \ve'$.

Towards constructing such a graph cover and a vector $\tve'$, we make the
following observations. Namely, fix some $d \in \Zpp$ and some $h_i \in \{
-1, +1 \}$, $i \in [d]$, and consider the hyperplane 
\begin{align*}
  \setA
    &\defeq
       \left\{
         \va \in \R^d
       \ \middle| \ 
         \sum\nolimits_{i \in [d]} h_i a_i = 0
       \right\}.
\end{align*}
Let $\va^{*} \in \setA$ be a vector with all its coordinates satisfying $-1
\leq a^{*}_i \leq +1$, $i \in [d]$. Let $\setA^{\square}$ be the set
\begin{align*}
  \setA^{\square}
    &\defeq
       \left\{
         \va \in \R^d
       \ \middle| \ 
         \begin{array}{ccc}
            a_i \in [0,+1] & \text{ if } & a^{*}_i > 0 \\
            a_i = 0        & \text{ if } & a^{*}_i = 0 \\
            a_i \in [-1,0] & \text{ if } & a^{*}_i < 0
         \end{array}
       \right\},
\end{align*}
which is a box around $\va^{*}$ whose vertices have only integer coordinates.

Consider now the set $\setA^{*} \defeq \setA \cap \setA^{\square}$, and let
$\setA'$ be the set of vertices of $\setA^{*}$. The set $\setA^{*}$ is a
polytope and, interestingly, it can be verified that the set of vertices of
$\setA^{*}$ is a subset of the set of vertices of $\setA^{\square}$, \ie, all
the points in $\setA'$ have integer coordinates. Because $\va^{*} \in
\setA^{*}$, this vector can be written as a convex combination of the vertices
of $\setA^{*}$, \ie, there are non-negative real numbers $\bigl\{ \beta_{\va'}
\bigr\}_{\va' \in \setA'}$ with $\sum_{\va' \in \setA'} \beta_{\va'} = 1$ such
that $\va^{*} = \sum_{\va' \in \setA'} \beta_{\va'} \va'$. Note that for all
$i \in [d]$ the following holds: if $a_i^{*} > 0$ then $a'_i \geq 0$ for all
$\va' \in \setA'$, if $a_i^{*} < 0$ then $a'_i \leq 0$ for all $\va' \in
\setA'$, and if $a_i^{*} = 0$ then $a'_i = 0$ for all $\va' \in \setA'$.

We now define $\mu \defeq \max_{i \in \setI} |e'_i|$ and apply the above
observations to our setup, in particular to the vector $\ve' / \mu$, whose
coordinates are rational numbers lying between $-1$ and $+1$
inclusive. Namely, for every $j \in \setJ$, we have $\sum_{i \in \setI_j}
h_{j,i} \cdot (e'_i / \mu) = 0$ with $h_{j,i} \in \{ -1, +1 \}$, $i \in
\setI_j$, and so there is a set $\setA'_j$ and non-negative rational numbers
$\bigl\{ \beta_{j, \va'_j} \bigr\}_{\va'_j \in \setA'_j}$ with $\sum_{\va'_j
  \in \setA'_j} \beta_{j, \va'_j} = 1$, such that $\ve'_{\setI_j} / \mu =
\sum_{i \in \setI} \beta_{j, \va'_j} \va'_j$ holds, where $\ve'_{\setI_j}$ is
the vector $\ve'$ restricted to the coordinates indexed by the set
$\setI_j$. Note that the set $\setA'_j$ is such that for all $i \in \setI_j$
the following holds: if $e'_i > 0$ then $a'_i \in \{ 0, +1 \}$ for all $\va'
\in \setA'_j$, if $e'_i < 0$ then $a'_i \in \{ -1, 0 \}$ for all $\va' \in
\setA'_j$, and if $e'_i = 0$ then $a'_i = 0$ for all $\va' \in \setA'_j$.

Let $\mu'$ be the largest positive real number such that $e'_i / \mu' \in
\Z$ for all $i \in \setI$ and such that $\beta_{j,\va'_j} / \mu' \in \Z$ for
all $j \in \setI$, $\va'_j \in \setA'_j$.

We are now ready to construct the promised $M$-fold cover of the base Tanner
graph and the valid configuration $\tve'$. We choose $M \defeq \mu / \mu'$
(clearly, $M \in \Zpp$), and so the constructed $\tve'$ will need to have the
properties shown in~\eqref{eq:CSRELzeroinf:graph:cover:assignment:1} with
$\gamma \defeq \mu / M = \mu'$. Without going into the details, the $M$-fold
cover with valid configuration $\tve'$ can be obtained with the help of the
above $\{ \beta_{j,\va'_j} \}_{j \in \setJ, \, \va'_j \in \setA'_j}$ values by
using a construction that is very similar to the explicit graph cover
construction in~\cite[Appendix~A.1]{Vontobel:Koetter:05:1:subm}. For example,
for every $i \in \setI$ with $e'_i > 0$ we set $M \cdot (e'_i / \mu) = e'_i /
\mu'$ of the values in $\bigl\{ \tilde e'_{(i,m)} \bigr\}_{m \in [M]}$ equal
to $\gamma$, and we set $M \cdot (1 - e'_i / \mu) = M - e'_i / \mu'$ of the
values in $\{ \tilde e'_{(i,m)} \}_{m \in [M]}$ equal to $0$, etc.. Similarly,
for every $j \in \setJ$ and $\va'_j \in \setA'_j$ we set the local
configuration of $M \cdot (\beta_{j,\va'_j} / \mu) = \beta_{j,\va'_j} / \mu'$
out of the $M$ copies of the $j$-th check node equal to $\va'_j$. Finally, the
edges between the variable and the constraint nodes of the $M$-fold cover of
the base Tanner graph are suitably defined. (Note that the definition of the
matrix in~\eqref{eq:CSRELzeroinf:constraint:1} implies that the edge
connections in the part of the graph cover corresponding to the right-hand
side of the matrix have already been pre-selected. However, this is not a
problem because the variable nodes associated with this part of the matrix
have degree one and because the above-mentioned constraint node assignments
can always be chosen suitably.)

This concludes the second part of the proof.

\end{document}